\documentclass[journal]{IEEEtran}

\usepackage{tikz}
\usepackage{multirow}
\usepackage{makecell}

\usepackage[utf8]{inputenc} 
\usepackage[T1]{fontenc}
\usepackage{url}
\usepackage{ifthen}
\usepackage{cite}
\usepackage[cmex10]{amsmath} 

\usepackage{bbm}                                                       
\usepackage{bm}
\usepackage{amsfonts,amssymb}
\usepackage{mathtools}
\usepackage{leftindex}

\usepackage{algorithm}
\usepackage[noend]{algpseudocode}

\usepackage[caption=false]{subfig}

\usepackage{comment}

\usepackage[normalem]{ulem}
\usepackage{soul}
\newcommand{\ignore}[1]{}

\DeclarePairedDelimiterX{\infdivx}[2]{(}{)}{%
  #1\;\delimsize\|\;#2%
}

\DeclarePairedDelimiter{\norm}{\lVert}{\rVert}

\newtheorem{claim}{Claim}

\newtheorem{lemma}{Lemma}

\newtheorem{remark}{Remark}
\newenvironment{proof}[1][Proof]{\noindent\textbf{#1.} }{\ \rule{0.5em}{0.5em}}

\DeclareMathOperator{\rect}{\square}

\DeclareMathOperator{\Var}{{Var}}

\DeclareMathOperator*{\argmin}{arg\,min}
\begin{document}
\title{Design of Threshold-Constrained  Indirect Quantizers}
\author{Ariel Doubchak, Tal Philosof, Uri Erez and Amit Berman

    \thanks{A.~Doubchak, T.~Philosof, and A.~Berman  are with 
     Samsung Semiconductor Israel Research and Development Center, Tel Aviv, Israel
    (e-mails: \texttt{\{ariel.d@samsung.com ,tal.philosof@samsung.com and amit.berman@samsung.com\}}).}
    \thanks{U.~Erez is with the School of Electrical Engineering, Tel Aviv University, Tel Aviv~6997801, Israel (e-mail: \texttt{\{uri@eng.tau.ac.il\}}).}
}

\maketitle


\begin{abstract}

We address the problem of indirect quantization of a source 
subject to a mean-squared error distortion constraint. 
A well-known result of Wolf and Ziv is that the problem can be reduced to  a standard (direct) quantization problem via a two-step approach: first apply the conditional expectation estimator, obtaining a ``new'' source, then solve for the optimal quantizer for the  latter source.
When quantization is implemented in hardware, however, invariably  constraints on the allowable class of quantizers are imposed, typically limiting the class to \emph{time-invariant} scalar quantizers with contiguous quantization cells.
In the present work, optimal indirect quantization subject to these constraints is considered. 
Necessary conditions an optimal quantizer within this class must satisfy are derived, in the form of generalized Lloyd-Max conditions, and an iterative algorithm for the design of such quantizers is proposed.
Furthermore, for the case of a scalar observation, we derive a non-iterative 
algorithm for finding the optimal indirect quantizer based on dynamic programming.

\end{abstract}

\section{Introduction}
\begin{figure}[htbp]
        \centering
        \includegraphics[width=\columnwidth]{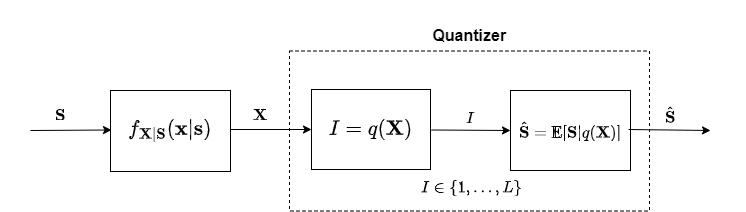}
        \caption{Indirect  quantization scenario subject to MSE distortion: An observation $X$, statistically dependent on an unobserved source $S$ is quantized, producing an index $I$, from which the reconstruction $S$ is produced.}
        \label{fig:single_phase}
    \end{figure}

The problem of indirect quantization \cite{dobrushin1962information,wolf1970transmission,witsenhausen1980indirect,sakrison1968source,berger1971rate,fine1965optimum,ephraim1988unified,gray1998quantization}, also referred to as remote source coding, is relevant to many practical settings where one does not have access to the source of interest, but rather observes measurements that are  statistically related  to the source. 
The goal is to quantize an observed $n$-dimensional vector of measurements $\mathbf{X}$  so as to allow reconstruction of a $k$-dimensional source vector $\mathbf{S}$ with minimal distortion as schematically depicted in Figure~\ref{fig:single_phase}. 
In this work, we will restrict attention to mean-square error (MSE) distortion.

The framework of indirect quantization is broad enough to encompass seemingly unrelated problems. 
On the one hand,
it can be applied as an immediate extension of the standard quantization problem to the case where one quantizes samples that have been corrupted by noise. In this case, the dimension of the observation vector is equal to that of the source vector, i.e. $n=k$, which is the usual regime in source coding.

On the other hand, the framework is also applicable to far less obvious generalizations. 
For example, the problem of Bayesian estimation of a parameter from a sequence of quantized observations, as studied, e.g., in \cite{stein2018performance,wen2015bayes,zeitler2012bayesian,stein2013quantization,mezghani2010multiple,bar2002doa,stein2016doa,liu2017one,yoffe2019direction}, 
can be cast under the indirect quantization framework, 
where the objective is to design a quantizer so as to minimize the estimation MSE.
In this scenario, the dimension of the observation vector will typically be much larger than that of the source (parameter) vector.
An example of this application appears in Section~\ref{sec:example}. 


We note that the problem of quantizer design with the goal of optimizing parameter estimation efficiency, for the asymptotic regime of $n \gg k$, was treated with different tools, by Stoica et al. \cite{cheng2022interval,gianelli2016one,stoica2021cramer} and in the context of distributed estimation, in
\cite{venkitasubramaniam2006score,venkitasubramaniam2007quantization,lam1993design,amari1995parameter}. 
Specifically, these works pursued minimization of metrics based on the Cram\'{e}r-Rao  bound and were geared towards the small-error region.\footnote{In general, large-error bounds, tend to be computationally complex and thus are less commonly used as a design metric; see, e.g., discussion in \cite{shalom2017efficient}.}


In the information-theoretic formulation of the indirect quantization problem, vector quantization is performed and the only constraint imposed on the quantizer is a rate constraint.
The optimal tradeoff between distortion and rate, that can be attained with quantization of arbitrarily large blocklength, is referred to as the indirect rate-distortion function. 
In many scenarios of interest, however, the quantizer must satisfy further constraints. 
See discussion of practical constraints in implementing hardware-limited indirect quantization in
\cite{shlezinger2019hardware,neuhaus2021task,bernardo2023design}. 
In particular, when quantization is performed as part of analog-to-digital conversion, invariably \emph{scalar} quantization is utilized. 

Furthermore, oftentimes, the relevant measure for the quantization ``rate'' is the (logarithm of the) number of thresholds needed for the implementation of the encoding function, rather than the (logarithm of the) cardinality of the index set. For example, this is the case in the read process in NAND flash memory where non-uniform scalar quantization is implemented via several voltage comparators, the number of which is a major resource one wishes to minimize; see, e.g., \cite{micheloni2010inside}. In order to formalize this distinction, let us introduce some notations.

\subsection{Problem Formulation}\label{sec:problem_formulation}




Consider the indirect quantization problem depicted in Figure~\ref{fig:single_phase}.
Denote the $k$-dimensional vector source of interest by $\mathbf{S}\in\mathbb{R}^k$, its joint probability density function (PDF) by $f_{\mathbf{S}}(\mathbf{s})$, and the $n$-dimensional observed measurement vector by $\mathbf{X}\in\mathbb{R}^n$. 
We assume that the distribution of the pair $(\mathbf{X},\mathbf{S})$ can be described by a joint probability density function $f_{\mathbf{SX}}(\mathbf{s},\mathbf{x})$. 
While the results derived in this work will be concerned, for the most part, with scalar quantization, we begin by defining the general setting.


A general vector quantizer (VQ) encoding function is a mapping
\begin{align}
{q}(\cdot):\mathbb{R}^n \rightarrow  \mathcal{I} = \{ 1, 2, \ldots, L \},
\label{eq:encoder_vq}
\end{align}
where $L$ is the number of output indices. 
Thus, $q(\mathbf{X})$ maps the observation vector to an index $\ell \in \mathcal{I}$.
We refer to $\log(L)/n$ as the rate of the quantizer. When no other constraints are imposed on the quantizer, we shall refer to the scenario as \emph{rate-constrained vector quantization} \cite{gersho2012vector,gray1998quantization}. 

The reconstruction of the source $\mathbf{S}$, given the quantization index, 
is given by the MMSE estimator, i.e., 
\begin{align}
    \hat{\mathbf{S}}=\mathbb{E}[\mathbf{S} \mid q(\mathbf{X})].
    \label{eq:quant_rec2}
\end{align}
The MSE associated with a quantizer $q(\cdot)$ is
\begin{align}
\text{MSE}_{q(\cdot)} =\mathbb{E}\left[\left\|\mathbf{S} - \hat{\mathbf{S}} \right\|^2\right].
\label{quant_MSE_gen}
\end{align}
We note that in the case where $\mathbf{X}=\mathbf{S}$, the problem reduces to the standard direct  quantization problem.

\begin{figure}[ht]
        \centering
        \includegraphics[width=0.6\columnwidth]     
        {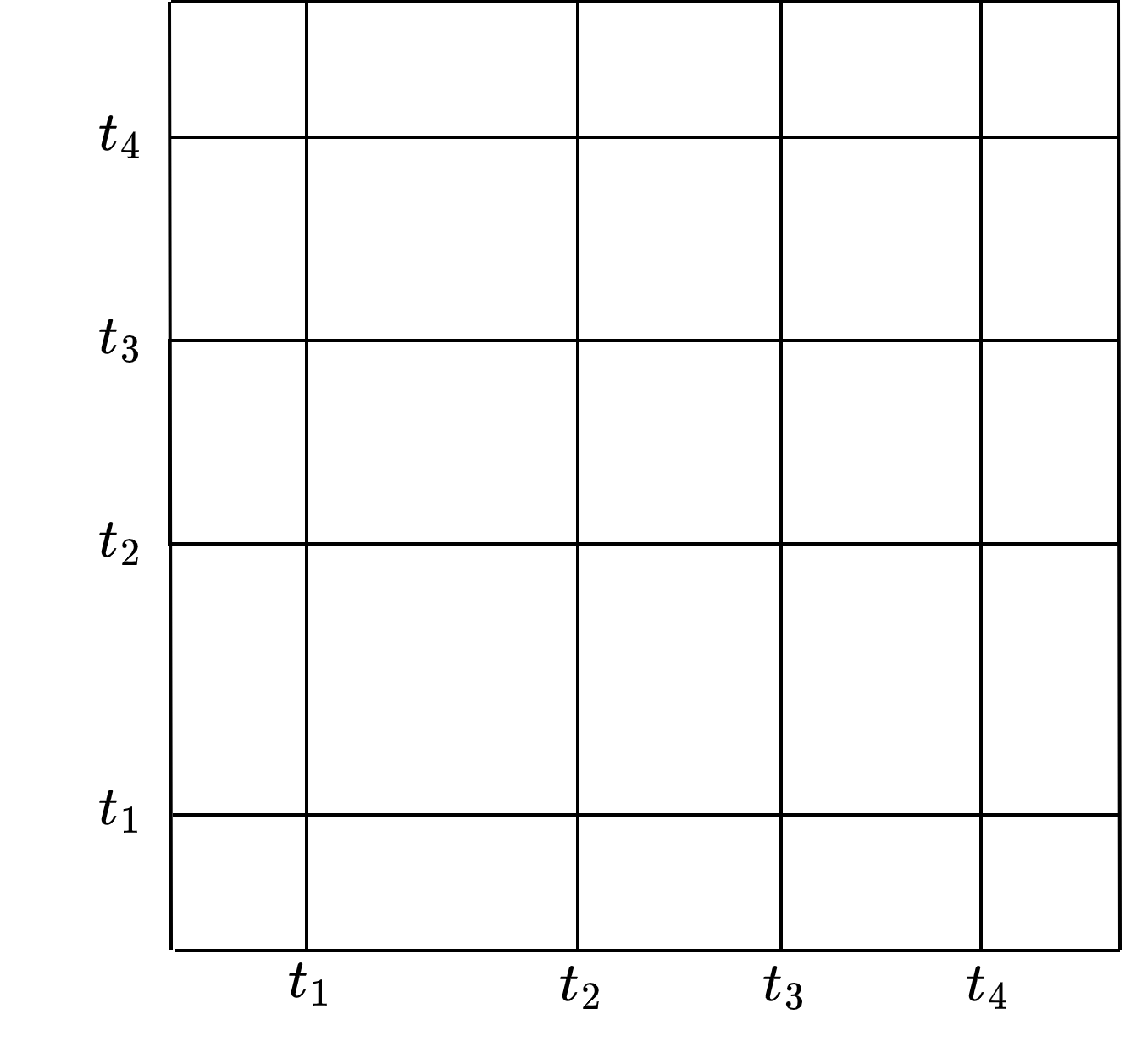}
        \caption{Induced quantization cells when scalar quantization applied to a two-dimensional ($n=2$) observation vector, with $T=4$ thresholds $t_1, t_2, t_3, t_4$.}
        \label{fig:2D_grid_intro}
    \end{figure}

A \emph{scalar} quantizer encoding function is specified by a scalar function 
\begin{align}
q(\cdot):\mathbb{R} \rightarrow  \mathcal{I} = \{ 1, 2, \ldots, L \}.
\label{eq:quant_enc_general}
\end{align}
We use the notation $q(\mathbf{X})$ to denote  component-wise application of the quantizer, i.e., 
\begin{align*}
    q(\mathbf{X})=(q(x_1),\ldots,q(x_n)).
\end{align*}
The application of a scalar quantizer to a vector observation is illustrated in Figure~\ref{fig:2D_grid_intro}.
When no further constraints are imposed, the scenario will be referred to as \emph{rate-constrained} scalar quantization.

\subsection{Rate-constrained vs. threshold-constrained scalar quantization}

Oftentimes, the relevant constraint is the number of thresholds needed for  implementation of the quantizer encoding function.
Specifically, this is the case when the quantizer is part of an analog-to-digital converter.
We will refer to such a scenario as \emph{threshold-constrained} scalar quantization. 

In this scenario, the scalar quantizer is constrained to have \emph{contiguous} quantization regions. 
Hence, the quantization function $q(\cdot)$ may be specified by $T = L-1$ thresholds $\{t_i\}_{i=1}^T$, where we set $t_0 = -\infty$ and $t_L= \infty$. Thus, subject to the constraint of contiguous cells, a quantizer encoding function is specified by a function    
\begin{align}
    q(x)=\ell, \, \text{if} \quad x \in (t_{\ell-1},t_\ell) \quad    , \, \ell=1,\ldots,L. \label{eq:quant_enc_contiguous}
\end{align}
One can then view (the logarithm of) $T+1$ as the relevant quantization ``rate".



\begin{figure}[htbp]
    \centering
    \subfloat[Rate-constraint quantizer ($T>L-1$) for $T=6$. Each of the four quantization cells, $L=4$, is denoted by a different color.]{\includegraphics[width=\linewidth]{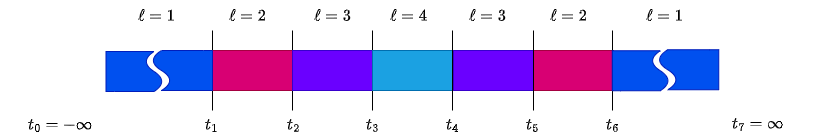}\label{fig:scalar_indirect_thr_rate}}
    \hfill
    \subfloat[Threshold-constrained quantizer {($T=L-1$)}. The six thresholds, $T=6$, induce seven contiguous quantization cells, $L=7$.]{\includegraphics[width=\linewidth]{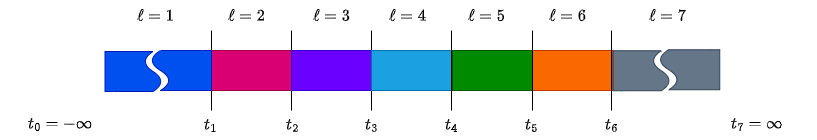}\label{fig:scalar_direct_thr_rate}}

    \caption{Rate-constrained vs. threshold-constrained scalar quantizer.} 
    \label{fig:rate_vs_thr_1D} 
\end{figure}

The distinction between the two scenarios is illustrated in Figure~\ref{fig:rate_vs_thr_1D}. Figure~\ref{fig:rate_vs_thr_1D}(a) depicts a scalar quantizer of general form (i.e., a rate-constrained quantizer),  where the encoder output is one of $L$ possible indices.  
Specifically, the number of cells is $L=4$ (represented by four colors) serves as the relevant ``rate'' in this scenario. 
Note that the number of associated thresholds is $T=6$ and that since not all quantization cells are contiguous, $T>L-1$. 
The implementation of the same quantizer under a threshold-constrained scenario is  illustrated in Figure~\ref{fig:rate_vs_thr_1D}(b) where now $T+1=7$ (represented by seven colors) is the relevant measure of ``rate''. 

We note that the distinction between the rate-constrained and threshold-constrained scenarios is of minor significance in the context of the direct quantization problem (with  MSE distortion measure), since the optimal quantization cells, when imposing a rate constraint, are always contiguous, and hence the two constraints are equivalent.\footnote{This follows since, for MSE distortion, the optimal cells are defined by the nearest-neighbor rule.} This is, however, \emph{not the case} in the indirect quantization problem as recounted below. 



\subsection{Overview of Previous Work}
The study of the remote quantization problem was initiated by the work of Dobrushin and Tsybakov \cite{dobrushin1962information}, and was subsequently extended by Sakrison \cite{sakrison1968source} and Wolf and Ziv in \cite{wolf1970transmission}, and by Berger \cite{berger1971rate}. The general formulation is due to Witsenhausen
\cite{witsenhausen1980indirect}. Algorithmic aspects of the quantizer design problem were addressed 
in the work of 
Fine \cite{fine1965optimum} where an iterative (Loyd-Max like) algorithm was derived. Sufficient conditions for convergence of the algorithm were derived by Ephraim and Gray \cite{ephraim1988unified},  for the case where the observations are the result of the source being corrupted by statistically-independent additive noise. All of these works considered the  rate-constrained scenario. For a comprehensive survey, the reader is referred to \cite{gray1998quantization}.

The results of the aforementioned works,  and specifically \cite{sakrison1968source,wolf1970transmission} in the context of MSE, state that the (rate-constrained) indirect  quantization problem may be reduced to a standard direct quantization problem via a two-step procedure: 
first apply the conditional expectation estimator, obtaining a ``virtual'' source, then solve for the optimal quantizer for the obtained virtual source, where by optimal we mean MSE-optimal for the direct quantization problem (of the virtual  source). 

\begin{figure}[htbp]
        \centering
        \includegraphics[width=\columnwidth]     
        {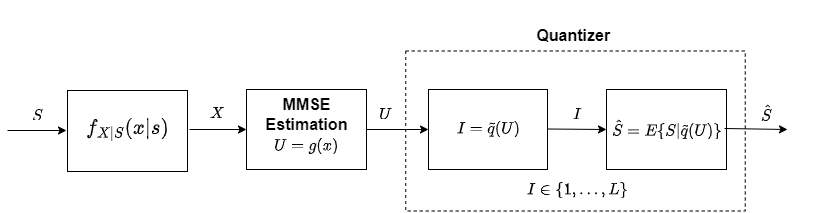}
        \caption{Two-step indirect quantization.}
        \label{fig:two_phase}
    \end{figure}

This result is illustrated by comparing Figures~\ref{fig:single_phase} and ~\ref{fig:two_phase}.
Figure~\ref{fig:single_phase} depicts a ``single-step'' implementation of an indirect quantizer. This structure directly follows from the problem formulation. One may alternatively attain the same performance via  a two-step approach as depicted in Figure~\ref{fig:two_phase} and described next. 

\subsubsection{Two-step approach and its limitations}
Let us define the optimal minimum mean- squared error (MMSE) estimator of the source from the observation by
\begin{align}
    g(x)\triangleq\mathbb{E}[S \mid X=x],
    \label{eq:mmse_function}
\end{align}
and define $U \triangleq g(X)$. 
Then one can design a quantization encoding function $\tilde{q}(\cdot)$, optimal in the MSE sense for quantizing $U$ viewed as a ``virtual'' source. In other words, $\tilde{q}(\cdot)$ is the optimal encoding function  for the direct quantization of $U$.\footnote{We note
that the two-step approach holds also for the case where the observation and 
source are vectors,
$\mathbf{X}$ and $\mathbf{S}$,
as appears  in \cite{wolf1970transmission}.}
Hence, this function will induce \emph{contiguous cells}.

The result of Wolf and Ziv \cite{wolf1970transmission}  states that the optimal quantization encoding function is obtained by the composition of $\tilde{q}$ and  $g(\cdot)$, i.e., 
\begin{align}
    q(x) = \tilde{q}(g(x)).
\end{align}

While the result is pleasing from a theoretical perspective, one must be cautious in its interpretation. 
Specifically, if $g(\cdot)$ is a non-monotone function, then the composition of $g(\cdot)$ and $\tilde{q}(\cdot)$ yields an encoding function $q(\cdot)$ where some cells are non-contiguous. Thus, the two-step approach is not beneficial in the threshold-constrained scenario since the constraint on the number of thresholds cannot be enforced in the domain of the virtual source $U$.
It follows that in the design problem of indirect quantizers, it is essential to treat the rate-constrained and threshold-constrained scenarios, separately. 



\subsubsection{Scalar vs. Vector Observations}
Another categorization of scenarios considered in prior art, as well as in the present work, is whether one quantized a scalar observation or a vector  observations. Some results apply only to the scalar case. In particular, it was shown by Bruce \cite{bruce1965optimum} that for the problem of direct quantization of a scalar observation (i.e., in this case, a scalar source), the optimal quantizer may be derived via dynamic programming (if one limits the set of possible threshold values to some finite resolution). 
Similarly, some of the results presented in the present paper  apply only to the problem of indirect quantization from a scalar observation, whereas others will generalize to the case  of quantization of a vector of observations, as we describe next.

\subsection{Contribution and Paper Organization}
In the present work, our goal is to:
\begin{itemize}
\item Explicitly state necessary conditions for optimality for the threshold-constrained indirect quantization problem. 
These conditions naturally lead to a corresponding iterative algorithm for the design of (in general suboptimal) practical quantizers for the problem, generalizing the Lloyd-Max algorithm. The results are complementary to those of Fine \cite{fine1965optimum} that addressed the problem in the absence of constraints on the structure of the quantization cells. 
The results are first presented in the context of a scalar observation, in 
Section~\ref{sec:scalar_iterative},
and are extended to the case of a vector of observations in Section~\ref{sec:indirect_vector_thr}.
\item
In Section~\ref{sec:indirect_scalar_opt_algo} we 
present results that apply only to the case of a scalar observation.
Specifically, we
extend the results of Bruce \cite{bruce1965optimum}  to the case of indirect quantization. We derive a dynamic-programming algorithm for the design of an optimal scalar quantizer for the  threshold-constrained scenario in Section~\ref{sec:indirect_scalar_algo_thr_opt}, and for the rate-constrained scenario in Section~\ref{sec:indirect_scalar_algo_rate_opt}.
\end{itemize}

{Table~\ref{tab:summary}} summarizes the state of knowledge and the related literature for all variants of the problem considered in this work, as well as our contributions. The table categorizes the problems according to the following delineations: direct/indirect quantization, scalar/vector observations, rate/threshold constraint, necessary conditions/optimal solution, and the state of our knowledge concerning the optimal solution.

\begin{table}[!htp]
    \begin{center}
    \begin{tabular}{|c|c||c||c|c|}
        \hline 
         \multicolumn{2}{|c||}{} & \textbf{Direct}  & \multicolumn{2}{c|}{\textbf{Indirect}}  \\
       \cline{3-5}
        \multicolumn{2}{|c||}{} & \makecell{Rate \\ Const. \\(fixed $L$)} & \makecell{Rate\\ Const.\\(fixed $L$)} & \makecell{Threshold \\ Const. \\(fixed $T$) }\\    
       \hline
       \hline
       \makecell{\textbf{Scalar} \\ \textbf{Observation} \\ $\mathbf{(n=1)}$} & \makecell{Necessary\\ Condition} & \makecell{Lloyd-Max \\ \cite{lloyd1982least,max1960quantizing}} & \makecell{Fine \cite{fine1965optimum} \\ Sec.~\ref{sec:fine}} &  Sec.~\ref{sec:extended_fine} \\
       \cline{2-5}
       & \makecell{Optimal\\ Solution} & \makecell{Bruce \cite{bruce1965optimum} \\ Sec.~\ref{sec:direct_scalar}} & Sec.~\ref{sec:indirect_scalar_algo_rate_opt} & Sec.~\ref{sec:indirect_scalar_algo_thr_opt}\\ 
        \hline 
       \makecell{\textbf{Vector} \\ \textbf{Observation} \\ $\mathbf{(n > 1)}$} & \makecell{Necessary \\ Condition} & \makecell{Gersho\\ \cite{Gersho82} } & \makecell{Sec.~\ref{sec:indirect_vector_rate} \\  Wolf-Ziv \\ \cite{wolf1970transmission}} & Sec.~\ref{sec:indirect_vector_thr}\\
       \cline{2-5}
       & \makecell{Optimal \\ Solution} & \makecell{Open \\ Problem} & \makecell{Open \\ Problem} & \makecell{Open \\ Problem}\\ 
       \hline 
    \end{tabular}
    \newline
    \caption{Summary of problems per category - scalar/vector quantization, direct/indirect quantization, rate/threshold constraints, necessary conditions/optimal solution.}
    \label{tab:summary}
    \end{center}
\end{table}

The rest of the paper is organized as follows. 
Section~\ref{sec:scalar_iterative} presents sufficient conditions for optimality of a scalar quantizer for the case of a scalar observation. 
Section~\ref{sec:indirect_scalar_opt_algo} derives a dynamic-programming algorithm for finding the optimal quantizer in the case of a scalar observation.
The results of Section~\ref{sec:scalar_iterative} are extended to the case of a vector of observations in Section~\ref{sec:extension}. 
The results are illustrated via numerical examples in Section~\ref{sec:example}. The paper concludes with a discussion in  Section~\ref{sec:discussion}. Technical derivations are relegated to the Appendix.

\section{Indirect Quantization: Necessary Conditions for Optimality, Scalar Observation}
\label{sec:scalar_iterative}

In this section, we present results for the scalar indirect quantization problem. We begin by recalling, in Section~\ref{sec:fine}, the  results of Fine \cite{fine1965optimum} for rate-constrained indirect scalar quantization. We the extend these results to the threshold-constrained scenario in Section~\ref{sec:extended_fine}, proving necessary conditions for optimality and deriving an associated iterative (Lloyd-Max like) algorithm for quantizer design for the latter setting. 





\subsection{Necessary Conditions for Optimality Subject to a Rate Constraint}
\label{sec:fine}

Fine's results are presented without proof in the following lemma. 

\begin{lemma}[Indirect SQ, Rate Constraint \cite{fine1965optimum}]\label{lem:Fine}
     Consider  indirect quantization of a source $S$ from an observation $X$, where the pair has joint PDF $f_{SX}(s,x)$. An optimal rate-constrained quantizer, as defined in \eqref{eq:quant_enc_general}  and \eqref{eq:quant_rec2}, i.e., one that minimizes $\mathbb{E}\left[(S-\hat{S})^2\right]$, must satisfy:
    \begin{itemize}
        \item Cell condition: Each quantization cell  \textcolor{black}{$\mathcal{R}_\ell,\;\ell=1,\ldots,L,$} satisfies
        \begin{align}
            \mathcal{R}_{\ell} = \left\{ x: \frac{\hat{s}_{\ell}+\hat{s}_{\ell-1}}{2}\leq g(x)\leq \frac{\hat{s}_{\ell}+\hat{s}_{\ell+1}}{2} \right\},\label{eq:thr_cond_Fine}
        \end{align}
        where
            $g(x)=\mathbb{E}[S \mid X=x]$, \textcolor{black}{and where we define $\hat{s}_0=-\infty$ and $\hat{s}_{L+1}=\infty$}
        \item Centroid condition:
         \begin{align}
            \hat{s}_{\ell} &= \mathbb{E}\left[S \mid x\in \mathcal{R}_{\ell}\right],\;\ell=1,\ldots,L.\label{eq:value_cond_Fine}
        \end{align}
    \end{itemize}
\end{lemma}

A consequence of Fine's optimality  conditions for indirect quantization is  that the quantization cells are, in general, no longer contiguous. Specifically, this can be inferred from the cell condition \eqref{eq:thr_cond_Fine}, since $g(x)$ can be a non-monotonic function. For this reason we use the terminology ``cell condition" rather than boundary condition.

\begin{remark}
Interestingly,  determining the quantizer cells based on $g(x)$, which represents the MMSE estimation of $S$ given $X=x$, as specified in \eqref{eq:thr_cond_Fine}, has a natural interpretation as a two-step approach. 
Namely, it corresponds to  setting boundaries according to the direct scalar quantization of the MMSE estimate $U=g(X)$.
This insight was leveraged to obtain the indirect rate-distortion function, subject to MSE distortion,  by Wolf and Ziv \cite{wolf1970transmission}.
It was subsequently extended beyond the case of MSE by Witsenhausen \cite{witsenhausen1980indirect}.
\end{remark}

\subsection{Necessary Conditions for Optimality Subject to a Thresholds Constraint}
\label{sec:extended_fine}

\textcolor{black}{
As discussed, in certain applications, particularly when quantization is implemented in hardware, the cost is better reflected in the number of thresholds used. In the case of scalar quantization, this constraint is equivalent to  requiring that the quantization cells  be contiguous.}

\begin{figure}[htbp]
        \centering
        \includegraphics[width=\columnwidth]     
        {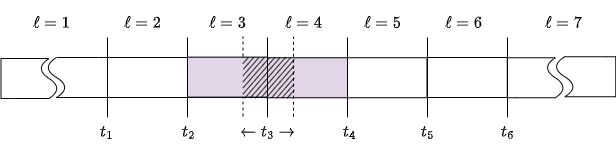}
        \caption{Necessary conditions for optimality for threshold-constrained scalar quantization: the two shaded (purple) cells are the ones affected when adjusting a single threshold ($t_3$).}
        \label{fig:1D_grid}
    \end{figure}

Specifically, restricting the number of quantization thresholds to $T$  implies that  the number of reconstruction values is $L = T+1$.\footnote{This follows since $L \leq T+1$, while increasing the number of quantization levels can only decrease the MSE. Consequently, the optimal quantization with minimal MSE must have contiguous cells.} 
In the following lemma, we provide necessary conditions for optimality for indirect scalar quantization for the   threshold-constrained scenario. The derivation of the conditions is qualitatively illustrated in Figure~\ref{fig:1D_grid}.

\begin{lemma}[Indirect SQ Subject to a Thresholds Constraint]\label{lem:general_lloyd_max}
     Consider  indirect quantization of a source $S$ from the scalar observation $X$, where the pair has joint PDF $f_{SX}(s,x)$. An optimal thresholds-constrained scalar quantizer, as defined in \eqref{eq:quant_enc_contiguous} and \eqref{eq:quant_rec2}, i.e., one that minimizes $\mathbb{E}\left[(S-\hat{S})\right]^2$, must satisfy:
    \begin{itemize}
        \item Boundary condition: Each threshold $t_\ell$, $\;\ell=1,\ldots,L-1,$ satisfies
        \begin{align}
             g(t_\ell) &= 
      \frac{\hat{s}_{\ell}+\hat{s}_{\ell+1}}{2},
             \label{eq:thr_cond}
            \end{align}
            where
            $g(t_\ell)=\mathbb{E}[S \mid X=t_\ell]$. 
        \item Centroid condition:
         \begin{align}
            \hat{s}_{\ell} &= \mathbb{E}\left[S \mid x\in (t_{\ell-1},t_{\ell} ) \right],\;\ell=1,\ldots,L, \label{eq:value_cond}
        \end{align}
            where $t_0=-\infty$ and $t_L= \infty$.
    \end{itemize}
\end{lemma}

\begin{remark}
    \begin{itemize}
        \item The centroid condition is identical in Lemmas~\ref{lem:Fine} and \ref{lem:general_lloyd_max} whereas the difference lies in the boundary conditions.
        \item Lemma \ref{lem:general_lloyd_max} generalizes the Lloyd-Max conditions for optimality, to the case of a source $S$ observed via a noisy observation $X$, namely to the scenario of indirect quantization. The conditions \eqref{eq:thr_cond} and \eqref{eq:value_cond} coincide with traditional Lloyd-Max conditions \cite{gersho2012vector} when setting $X=S$.
    \end{itemize}
\end{remark}

\begin{proof}
Consider a given set of thresholds \mbox{$\mathbf{t}=[t_0 = -\infty, t_1,\ldots,t_{L-1}, t_L = \infty]$} and a set of quantization reconstruction values, i.e.,  $\hat{s}_{\ell},\ell=1,\ldots,L$. 

The MSE \eqref{quant_MSE_gen}, in the case of scalar quantization,  is explicitly given by 
\begin{align}
    \text{MSE}_{\mathbf{t}} &= \sum_{\ell=1}^L \int_{s \in \mathbb{R}} \int_{t_{\ell-1}}^{t_{\ell}}(s-\hat{s}_{\ell})^2f_{SX}(s,x)dx ds.
\end{align}

Consider now the effect on the MSE when adjusting a single threshold value as depicted in   Figure~\ref{fig:1D_grid}.
Following the Leibniz integral rule, the partial derivative of the MSE with respect to a threshold $t_{\ell}$ (holding all other parameters fixed) is
 \begin{align}
     B(t_\ell) & \triangleq \frac{\partial \text{MSE}_\mathbf{t}}{\partial t_{\ell}} \\ 
     %
     &= \int_{s \in \mathbb{R}} f_{SX}(s,t_{\ell})\left\{(s-\hat{s}_{\ell})^2 - (s-\hat{s}_{\ell+1})^2\right\}ds\\
     &= \int_{s \in \mathbb{R}} f_{SX}(s,t_{\ell})(\hat{s}_{\ell}-\hat{s}_{\ell+1})(\hat{s}_{\ell}+\hat{s}_{\ell+1}-2s)ds.\label{eq:der_mse_t}
 \end{align}
Similarly, consider the effect on the MSE of adjusting a reconstruction value. 
The partial  derivative of the MSE with respect to a reconstruction value $\hat{s}_{\ell}$ is given by
\begin{align}
    C(\hat{s}_\ell)   \triangleq \frac{\partial \text{MSE}_\mathbf{t}}{\partial \hat{s}_{\ell}} 
     &= -2\int_{t_{\ell-1}}^{t_{\ell}} \int_{s \in \mathbb{R}} (s-\hat{s}_{\ell}) f_{SX}(s,x)dxds.\label{eq:der_mse_s}
 \end{align}
 
Necessary conditions for  optimality for threshold values $t_{\ell}$ are that 
\begin{align}
    B(t_\ell)=0,
     \label{eq:simple_boundary1}
\end{align}
for $\ell=1,\ldots,L-1$. Therefore, from \eqref{eq:der_mse_t}, we get that
 \begin{align}
    B(t_\ell)= \int_{s \in \mathbb{R}} f_{SX}(s,t_{\ell})(\hat{s}_{\ell}+\hat{s}_{\ell+1}-2s)ds = 0.
    \label{eq:simple_boundary}
 \end{align}
This condition can be rewritten as follows 
 \begin{align}
 \frac{\hat{s}_{\ell}+\hat{s}_{\ell+1}}{2}  
     &= \frac{\int_{s \in \mathbb{R}} sf_{SX}(s,t_{\ell}) ds}{\int_{s \in \mathbb{R}} f_{SX}(s,t_{\ell}) ds}\\
     &= \int_{s \in \mathbb{R}} \frac{sf_{SX}(s,t_{\ell})}{f_{X}(t_{\ell})} ds\label{eq:cond_t_eqn100}\\
     &= \int_{s \in \mathbb{R}} sf_{S|X}(s|t_{\ell})ds\label{eq:cond_t_eqn110}\\
     &= \mathbb{E}(S|X = t_{\ell}),
 \end{align}
where \eqref{eq:cond_t_eqn100} follows since $f_{X}(t_{\ell}) = \int_{s \in \mathbb{R}} f_{SX}(s,t_{\ell})ds$ and \eqref{eq:cond_t_eqn110} follows by Bayes' rule. This proves the boundary conditions \eqref{eq:thr_cond}.



The centroid condition is simply the MMSE estimator for the source given the interval index. 
Thus, we have established the necessary conditions \eqref{eq:thr_cond} and \eqref{eq:value_cond}.
\end{proof}

\subsubsection{Iterative Algorithm for Threshold-Constrained}\label{sec:indirect_scalar_iter_thr_algo}
The boundary and the centroid conditions, \eqref{eq:thr_cond} and \eqref{eq:value_cond},  specified in Lemma~\ref{lem:general_lloyd_max}, naturally lead to an iterative algorithm for setting the thresholds and  reconstruction levels of the quantization cells. 

The algorithm initializes the thresholds $\mathbf{t}$ arbitrarily and computes a set of reconstruction values according to  \eqref{eq:value_cond}. Subsequently, a new set of thresholds $\mathbf{t}$ is evaluated based on \eqref{eq:thr_cond}. The algorithm proceeds iteratively, evaluating a set of reconstruction values and a set of thresholds until it converges to some desired accuracy ($\varepsilon$). The algorithm is summarized in Algorithm~\ref{algo:indirect_LM}.


\begin{algorithm}
\caption{Iterative algorithm for indirect scalar quantization subject to threshold-constrained}

\begin{algorithmic}[1]

\Procedure{Indirect\_Quantization}{$f(s),f(s,x)$}       
    \State Set $n = 1$ 
    \State Set $\mathbf{t}^{(n)}$ arbitrarily
    \State Set $\text{MSE}^{(n-1)} = \infty$
    \State Set $\text{MSE}^{(n)} = 0$

    \While{$ \text{MSE}^{(n-1)} - \text{MSE}^{(n)} > \varepsilon$}  
        \State Set $\hat{s}^{(n)}_{\ell} = \mathbb{E}\left(S|X\in[t^{(n)}_{\ell-1},t^{(n)}_{\ell}]\right),\, \ell=[1,L-1]$
        \State Set $g\left(t^{(n)}_\ell\right) = 
     \frac{\hat{s}^{(n)}_{\ell}+\hat{s}^{(n)}_{\ell+1}}{2},\quad\ell=[1,L-1]$
        \State Solve $t^{(n)}_{\ell}$: $g\left(t^{(n)}_\ell\right)=\mathbb{E}[S \mid X=t_\ell]$
        \State Set $n=n+1$
        \State ${\displaystyle    \text{MSE}^{(n)} = \sum_{\ell=1}^L \int_{s \in \mathbb{R}} (s-\hat{s}^{(n)}_{\ell})^2 f_{S \mid X\in(t^{(n)}_{\ell-1},t^{(n)}_{\ell})}(s)ds}$
    \EndWhile  
\EndProcedure

\end{algorithmic}
\label{algo:indirect_LM}
\end{algorithm}

\section{Indirect Quantization: Optimal Solution for a Scalar Observation via Dynamic Programming}
\label{sec:indirect_scalar_opt_algo}

The Lloyd-Max conditions for optimality for the direct quantization problem, as well as the associated iterative algorithm for quantizer design, are well known \cite{gersho2012vector}. 

The \emph{optimal} setting of thresholds for direct scalar quantization was derived in \cite{bruce1965optimum} by applying dynamic programming. As these  results are not as well known, we begin by recalling them in Section~\ref{sec:direct_scalar}. 
The results are then extended to the indirect quantization problem subject to a rate constraint in Section~\ref{sec:indirect_scalar_algo_rate_opt} and to the threshold-constrained scenario in Section~\ref{sec:indirect_scalar_algo_thr_opt}.

\subsection{Optimal Quantizer in the Rate-Constrained Scenario}
\subsubsection{Background: Dynamic Programming Solution for Direct Quantization (Bruce \cite{bruce1965optimum})}\label{sec:direct_scalar}


In many practical applications, threshold settings are often constrained to be chosen from a pre-determined finite set. Typically, these values belong to some fine-resolution grid. This limitation arises from the finite accuracy with which threshold values can be set. For instance, in the case of a voltage axis, the threshold voltage cannot be determined with infinite accuracy. In such cases, one may assume that the thresholds can take on one of $M$ values, where typically $M \gg L$.  We define the set of $M$ values as $\mathcal{T}$, i.e., $|\mathcal{T}| = M$.

The MSE for any set of thresholds $\mathbf{t}$ is given by
\begin{align}
    \text{MSE}_{\mathbf{t}} &= \sum_{\ell=1}^L  \int_{t_{\ell-1}}^{t_{\ell}}(s-\hat{s}_{\ell})^2f_{S}(s) ds
\end{align}
where for any $\mathbf{t}$, the optimal reconstruction values are $\hat{s}_{\ell} = \mathbb{E}[S|S\in [t_{\ell-1},t_{\ell}]]$. This problem can be written as the following minimization problem: 
\begin{equation}
\begin{aligned}
     \mathbf{t}^o = \argmin_{\mathbf{t}\in\mathcal{T}^{L-1}}\text{MSE}_{\mathbf{t}}
     = \argmin_{\mathbf{t}\in\mathcal{T}^{L-1}} \sum_{\ell=1}^L  \int_{t_{\ell-1}}^{t_{\ell}}(s-\hat{s}_{\ell})^2f_{S}(s) ds.     
\label{eq:direct_scalar_opt}
 \end{aligned}
 \end{equation}
where $t_0=-\infty$ and $t_{L}=\infty$, and $\mathcal{T}^{L-1}$ is the $L-1$ dimensional Cartesian product of $\mathcal{T}$. 

Bruce \cite{bruce1965optimum} developed a dynamic-programming algorithm to solve \eqref{eq:direct_scalar_opt}. 
In contrast to the Lloyd-Max iterative algorithm \cite{max1960quantizing, lloyd1982least}, which converges to a local minimum, Bruce's algorithm is a non-iterative approach that achieves the optimal solution.


Following \cite{bruce1965optimum}, the inner integral in \eqref{eq:direct_scalar_opt} can be defined as a ``partial MSE''
\begin{equation}
\begin{aligned}
\text{mse}(t_{\ell-1},t_{\ell})&\triangleq \int_{t_{\ell-1}}^{t_{\ell}}(s-\hat{s}_{\ell})^2f_{S}(s) ds.     
\end{aligned}
\end{equation}

We use the following definition for a \textit{state}, which allows formulating the optimization problem as dynamic programming:
\begin{align}
    {\sf{S}}(t_{\ell}) \triangleq \min_{t_{\ell-1}: t_{\ell-1} <t_{\ell}}\Big\{{\sf{S}}(t_{{\ell}-1})+\text{mse}(t_{{\ell}-1},t_{\ell})\Big\},\;{\ell}=1,\ldots,L,\label{eq:scalar_direct_DP_state}
\end{align}
where ${\sf{S}}(t_0) = 0$. The optimization problem in \eqref{eq:direct_scalar_opt} can be expressed recursively using \eqref{eq:scalar_direct_DP_state} as follows:
\begin{equation}
\begin{aligned}
    &\min_{\mathbf{t}}\;\text{MSE}_\mathbf{t} = \min_{\mathbf{t}}\sum_{\ell=1}^{L} \text{mse}(t_{\ell-1},t_{\ell})\\
    &= \min_{t_{L-1},\ldots,t_{1}}\Big(\text{mse}(t_{0},t_{1})+\text{mse}(t_{1},t_{2})+\ldots+\text{mse}(t_{L-1},t_{L})\Big)\\
    &= \min_{t_{L-1},\ldots,t_{1}}\Big({\sf{S}}(t_{1})+\text{mse}(t_{1},t_{2})+\ldots+\text{mse}(t_{L-1},t_{L})\Big) \\
    &= \min_{t_{L-1},\ldots,t_{2}}\Big({\sf{S}}(t_{2})+ \text{mse}(t_{2},t_{3})\ldots+\text{mse}(t_{L-1},t_{L})\Big). 
\end{aligned}\label{eq:scalar_direct_DP_der1} 
\end{equation}
Continuing recursively, we obtain: 
\begin{equation}
\begin{aligned}
    \min_{\mathbf{t}}&\;\text{MSE}_\mathbf{t}  
    = \min_{t_{L-1}}\Big\{ {\sf{S}}(t_{L-1})+\text{mse}(t_{L-1},t_{L})\Big\}
    = {\sf{S}}(t_L).
\end{aligned}\label{eq:scalar_direct_DP_der2} 
\end{equation}

The computation in \eqref{eq:scalar_direct_DP_der1} and \eqref{eq:scalar_direct_DP_der2} employs a recursive technique to find the optimal solution for $\mathbf{t}$, known as the dynamic programming algorithm. This optimization process is broken down into a sequence of partial minimization problems. The algorithm starts with cost ${\sf{S}}(t_0)=0$ at stage $0$, searching for an optimal $t_1$ among all feasible options of $t_1$. Its objective is to minimize ${\sf{S}}(t_1) + \text{mse}(t_1, t_2)$ for any $t_1$ satisfying $t_1 < t_2$. Subsequently, at each stage, the algorithm identifies the optimal $t_{\ell}$ such that ${\sf{S}}(t_{\ell}) + \text{mse}(t_{\ell}, t_{\ell+1})$ is minimized for any $t_{\ell}$ satisfying $t_{\ell} < t_{\ell+1}$, continuing through $L$ stages. Finally, the last stage corresponds to the minimum state ${\sf{S}}(t_L)$. The selected states along the sequence with minimal ${\sf{S}}(t_L)$ yield the optimal thresholds $\mathbf{t}^o$ as defined in \eqref{eq:direct_scalar_opt}.




\subsubsection{Dynamic Programming Solution for Indirect Quantization Subject to a Rate Constraint}
\label{sec:indirect_scalar_algo_rate_opt}
{
We now describe an extension of the dynamic-programming algorithm derived in \cite{bruce1965optimum} to the problem of indirect scalar quantization subject to a rate constraint. 

The necessary conditions for optimality for indirect scalar quantization subject to a rate constraint, as derived in \cite{fine1965optimum}, imply that the optimal quantizer cells are expressed via thresholds in terms of the function  $g(x)=\mathbb{E}[S \mid X=x]$. Specifically, this property stems from the cell condition \eqref{eq:thr_cond_Fine} in Lemma~\ref{lem:Fine}.

Accordingly, to derive the dynamic-programming algorithm, we introduce the random variable $$U=g(X)=\mathbb{E}[S \mid X].$$ 
We denote the thresholds for $U$ by  $\mathbf{v} = [v_1,v_2,\ldots,v_{L-1}]$, which define $L$ reconstruction values. Here, $v_0=-\infty$, $v_L=\infty$, and $v_{\ell} < v_{\ell+1}$.

The MSE may explicitly be  written as:
\begin{align}
\text{MSE}_{\mathbf{v}} &= \sum_{\ell=1}^L \int_{v_{\ell-1}}^{v_{\ell}} \int_{s \in \mathbb{R}} (s-\hat{s}_{\ell})^2 f_{SU}(s,u)du ds,
\end{align}
where $\hat{s}_{\ell} = \mathbb{E}[S|U\in [v_{\ell-1},v_{\ell}]]$ for any set of thresholds $\mathbf{v}$, by the centroid condition \eqref{eq:value_cond_Fine}.


As in the  case of direct quantization \eqref{eq:direct_scalar_opt}, the thresholds for $X$ are restricted to belong to a set of finite resolution.  This limits the thresholds for $U=g(X)$ to a set of finite resolution as well which we denote by $\mathcal{V}$, where $|\mathcal{V}| = N$.
The task of finding the optimal thresholds can thus be formulated as the following minimization problem:
\begin{equation}
\begin{aligned}
     \mathbf{v}^o &= \argmin_{\mathbf{v}\in\mathcal{V}^{L-1}}\text{MSE}_{\mathbf{v}}\\
     &= \argmin_{\mathbf{v}\in\mathcal{V}^{L-1}} \sum_{\ell=1}^L  \int_{v_{\ell-1}}^{v_{\ell}} \int_{s \in \mathbb{R}} (s-\hat{s}_{\ell})^2f_{SU}(s,u)du ds,  
\label{eq:indirect_scalar_opt_rate}
 \end{aligned}
 \end{equation}
where $v_0=-\infty$ and $v_{L}=\infty$. We denote the inner integrals in \eqref{eq:indirect_scalar_opt_rate} by
\begin{align}
    \text{mse}(v_{\ell-1},v_{\ell})&\triangleq \int_{v_{\ell-1}}^{v_{\ell}} \int_{s \in \mathbb{R}} (s-\hat{s}_{\ell})^2 f_{SX}(s,x)dx ds. \label{eq:indirect_scalar_rate_metric_def}
\end{align}

In order to present the optimization problem  \eqref{eq:indirect_scalar_opt_rate} in dynamic-programming formalism, we use the following definition of state:
\begin{align}
    {\sf{S}}(v_{\ell}) \triangleq \min_{v_{\ell-1}: v_{\ell-1} <v_{\ell}}\Big\{{\sf{S}}(v_{{\ell}-1})+\text{mse}(v_{{\ell}-1},v_{\ell})\Big\},\;{\ell}=1,\ldots,L,\label{eq:indirect_scalar_rate_DP_state_def}
\end{align}
where ${\sf{S}}(t_0) = 0$. Similar to the derivation in \eqref{eq:scalar_direct_DP_der1} and \eqref{eq:scalar_direct_DP_der2}, the optimization problem in \eqref{eq:indirect_scalar_opt_rate} can be expressed recursively using \eqref{eq:indirect_scalar_rate_DP_state_def}. Specifically, we have:
\begin{equation}
\begin{aligned}
    \max_{\mathbf{v}}\;\text{MSE}_{\mathbf{v}} &= \min_{\mathbf{v}}\sum_{\ell=1}^{L} \text{mse}(v_{\ell-1},v_{\ell})\\
    &= \min_{v_{L-1},\ldots,v_{\ell}}\Big({\sf{S}}(v_{\ell})+ \sum_{j=\ell}^{L-1}\text{mse}(v_{j},v_{j+1})\Big) \\
    &= \min_{v_{L-1}}\Big\{ {\sf{S}}(v_{L-1})+\text{mse}(v_{L-1},v_{L})\Big\}\\
    &= {\sf{S}}(v_L).
\end{aligned}\label{eq:scalar_indirect_rate_DP_der} 
\end{equation}

The computation in \eqref{eq:scalar_indirect_rate_DP_der} yields a dynamic programming algorithm, summarized in Algorithm~\ref{algo:indirect_scalar_rate_DP}. The metric \eqref{eq:indirect_scalar_rate_metric_def}, needs to be evaluated for all $t_i < t_j$, resulting in $N(N-1)/2$ feasible values. The metric \eqref{eq:indirect_scalar_rate_metric_def} can be reformulated as:
\begin{align}
    \text{mse}(v_i,v_j) = \Pr(U\in[v_i,v_j])\Var\left(S|U\in[v_i,v_j]\right),\label{eq:indirect_scalar_rate_metric}     
\end{align}



\begin{algorithm}
\caption{Optimal solution for indirect scalar quantization subject to rate-constrained}

\begin{algorithmic}[1]
\State \textbf{Input:} $f_{SX}(s,x), g(X)$.
\State Compute $f_{SU}(s,u)$, where $U = g(X)$. 
\State Compute from \eqref{eq:indirect_scalar_rate_metric} all metrics: $\text{mse}(v_s,v_f),\;\forall \, v_s, v_f\in\mathcal{V}$ and $v_s < v_f$.
\State Set $S(v_0) = 0,v_0 = -\infty, v_L = \infty$.
\State For $\ell = 1,\ldots,L$: compute $S(v_{\ell})$ from \eqref{eq:indirect_scalar_rate_DP_state_def} for $\forall v_{\ell}\in\mathcal{T}$. 
\State Trace back from ${\sf{S}}(v_L)$ to ${\sf{S}}(v_0)$ and extract the optimal thresholds in each step of minimization- $(v_{L-1}^o,v_{L-2}^o, \ldots, v_1^o)$.
\State Set $\text{MSE} = {\sf{S}}(v_L)$.
\State \textbf{Output:} $\text{MSE}$, $(v_{L-1}^o,v_{L-2}^o, \ldots, v_1^o)$
\end{algorithmic}
\label{algo:indirect_scalar_rate_DP}
\end{algorithm}


It is important to note that Algorithm~\ref{algo:indirect_scalar_rate_DP} seeks the optimal locations of $L-1$ thresholds in the $U$ domain (where $U=g(X)$), minimizing the MSE and yielding $L$ quantization reconstruction levels. However, since the observations are in the $X$ domain, the $L$ quantization intervals in $U$ must be converted into intervals in $X$. This translation is achieved through a fixed mapping, that may be implemented  via a lookup table. Generally, each interval in $U$ may correspond to several intervals in $X$. 

\begin{remark}
    An alternative approach to avoid the explicit computation of the joint distribution $f_{SU}(s,u)$ and the subsequent calculation of $\text{Var}\left(S|U\in[v_i,v_j]\right)$ involves sorting the $N$ values (corresponding to the finite-resolution grid) of $g(x)$, in ascending order, 
    with the corresponding order of $X$ denoted as $\tilde{X}$. Then, we can substitute $\tilde{X}$ for $U$ in the dynamic programming algorithm.
\end{remark}


\subsection{Optimal Quantizer in the Threshold-Constrained Scenario}
\label{sec:indirect_scalar_algo_thr_opt} 
The extension of the dynamic-programming algorithm derived in \cite{bruce1965optimum} to the problem of threshold-constrained indirect quantization follows by writing \eqref{quant_MSE_gen} explicitly. The quantization MSE can be written as: 
\begin{align}
    \text{MSE}_{\mathbf{t}} &= \sum_{\ell=1}^L \int_{t_{\ell-1}}^{t_{\ell}} \int_{s \in \mathbb{R}} (s-\hat{s}_{\ell})^2f_{SX}(s,x)dx ds,
\end{align}
where for any set of thresholds $\mathbf{t}$, the optimal reconstruction values are \mbox{$\hat{s}_{\ell} = \mathbb{E}[S|S\in [t_{\ell-1},t_{\ell}]]$} by the centroid condition  \eqref{eq:value_cond}. 

We again assume that the set of possible thresholds 
is a discrete set of values $\mathcal{T}$
and thus the optimization can be formulated as the following minimization problem:
\begin{equation}
\begin{aligned}
     \mathbf{t}^o &= \argmin_{\mathbf{t}\in\mathcal{T}^{L-1}}\text{MSE}_{\mathbf{t}}\\
     &= \argmin_{\mathbf{t}\in\mathcal{T}^{L-1}} \sum_{\ell=1}^L  \int_{t_{\ell-1}}^{t_{\ell}} \int_{s \in \mathbb{R}} (s-\hat{s}_{\ell})^2f_{SX}(s,x)dx ds.     
\label{eq:indirect_scalar_opt}
 \end{aligned}
 \end{equation}
where $t_0=-\infty$ and $t_{L}=\infty$. We denote the inner integrals in \eqref{eq:indirect_scalar_opt} by 
\begin{align}
    \text{mse}(t_{\ell-1},t_{\ell})&\triangleq \int_{t_{\ell-1}}^{t_{\ell}} \int_{s \in \mathbb{R}} (s-\hat{s}_{\ell})^2f_{SX}(s,x)dx ds.   
\label{eq:indirect_scalar_metric}
\end{align}

Clearly, the problem in \eqref{eq:indirect_scalar_opt} is similar to \eqref{eq:indirect_scalar_opt_rate}, with $X$ and $\mathbf{t}$ substituting $U$ and $\mathbf{v}$, respectively. Consequently,
we obtain a dynamic-programming algorithm where the metric and the states are defined as follows:
\begin{align}
    \text{mse}(t_{i},t_{j}) &\triangleq \int_{t_{i}}^{t_{j}} \int_{s \in \mathbb{R}} (s-\hat{s}_{i})^2f_{SX}(s,x)dx ds \nonumber\\
    &= \Pr(X\in[t_i,t_j])\Var\left(S|X\in[t_i,t_j]\right),\label{eq:indirect_scalar_thr_metric} 
\end{align}
and 
\begin{align}
{\sf{S}}(t_{\ell}) \triangleq \min_{t_{\ell-1}: t_{\ell-1} <t_{\ell}}\Big\{{\sf{S}}(t_{{\ell}-1})+\text{mse}(t_{{\ell}-1},t_{\ell})\Big\},\label{eq:indirect_scalar_thr_DP_state_def}
\end{align}
for $\ell = 1,\ldots,L-1$ and ${\sf{S}}(t_0) = 0$.

The dynamic programming algorithm is summarized in Algorithm~\ref{algo:indirect_scalar_thr_DP}.


\begin{algorithm}
\caption{Optimal solution for indirect scalar quantization subject to threshold-constrained}

\begin{algorithmic}[1]
\State \textbf{Input:} $f_{SX}(s,x)$.
\State Compute from \eqref{eq:indirect_scalar_thr_metric} all metrics: $\text{mse}(t_s,t_f),\;\forall t_s, t_f\in\mathcal{T}$ and $t_s < t_f$.
\State Set ${\sf{S}}(t_0) = 0,t_0 = -\infty, t_L = \infty$.
\State For $\ell = 1,\ldots,L$: compute ${\sf{S}}(t_{\ell})$ from \eqref{eq:indirect_scalar_thr_DP_state_def} for $\forall t_{\ell}\in\mathcal{T}$. 
\State Trace back from ${\sf{S}}(t_L)$ to ${\sf{S}}(t_0)$ and extract the optimal thresholds in each step of minimization- $(t_{L-1}^o,t_{L-2}^o, \ldots, t_1^o)$.
\State Set $\text{MSE} = {\sf{S}}(t_L)$.
\State \textbf{Output:} $\text{MSE}$, $(t_{L-1}^o,t_{L-2}^o, \ldots, t_1^o)$
\end{algorithmic}
\label{algo:indirect_scalar_thr_DP}
\end{algorithm}

}


\section{Indirect Vector Case: Necessary Conditions for Optimality and Iterative Algorithm}
\label{sec:extension}

We now return to the general problem formulation as presented in Section~\ref{sec:problem_formulation} where a $k$-dimensional source is to reproduced based on the quantization of an $n$-dimensional  observation
vector. 

For sake of completeness, we first discuss the rate-constrained scenario where  \emph{general vector} quantization of the observations is allowed. 
Specifically, we recall in  Section~\ref{sec:indirect_vector_rate}
how
the necessary conditions for optimality presented  in Section~\ref{sec:fine} for a scalar observation were generalized to rate-constrained vector quantization by Wolf and Ziv \cite{wolf1970transmission}.

Then, in Section~\ref{sec:indirect_vector_thr}, we turn to consider the case where threshold-constrained \emph{scalar} quantization is applied to the observation vector, which is the scenario of interest in the present work. We derive necessary conditions for optimality for this setting. 

\subsection{Necessary Conditions for Optimality: Rate Constraint}\label{sec:indirect_vector_rate}

Consider a general partition of $\mathbb{R}^n$ into $L$ cells, namely,  $\mathcal{R}_{\ell}, {\ell}=1,\ldots,L$, such that $\bigcup_{\ell=1}^L\mathcal{R}_{\ell} = \mathbb{R}^n$, where $\mathcal{R}_\ell\bigcap\mathcal{R}_{\ell'} = \emptyset$ for $\ell' \neq \ell$, and  denote the  reconstruction vectors for each cell by $\mathbf{\hat{s}}_{\ell}$.  
Using these  regions as quantization cells for the encoding function $q(\cdot)$, and the  reconstruction vectors $\mathbf{\hat{s}}_{\ell}$ for the decoding function, the MSE \eqref{quant_MSE_gen} can be written as
\begin{align}
 \text{MSE}
 &= \sum_{\ell=1}^L \int_{s \in \mathbb{R}} \int_{\mathcal{R}_{\ell}}\|\mathbf{s}-\mathbf{\hat{s}}_{\ell}\|^2 f_{\mathbf{SX}}(\mathbf{s},\mathbf{x})\mathbf{dx ds}.
\label{eq:MSE_VQ_general}
\end{align}

The necessary conditions for optimality for indirect vector quantization subject to a rate constraint were derived by Wolf and Ziv in \cite{wolf1970transmission} and are a natural generalization of the solution presented for the scalar case in Section~\ref{sec:fine}. The optimal solution is attained by applying optimal vector quantization to a transformed source, the latter defined as the MMSE estimation of $\mathbf{S}$ from the observation $\mathbf{X}$. Namely, one specifies the optimal quantization condition for a virtual (transformed) source 

\textcolor{black}{
$$ \mathbf{U} \triangleq \mathbf{g}(\mathbf{X})=\mathbb{E}[\mathbf{S} \mid \mathbf{X} = \mathbf{x}].$$}
In the following claim, we summarize the 
resulting
necessary conditions for optimality in indirect vector quantization subject only to a rate constraint.

\begin{claim}[Indirect VQ, Rate Constraint]\label{cor:vec_lloyd_max}
Consider  indirect quantization of a vector source $\mathbf{S}$ from the observation $\mathbf{X}$, where the pair has joint PDF $f_{\mathbf{SX}}(\mathbf{s},\mathbf{x})$. An optimal quantizer, i.e., one that minimizes $\mathbb{E}\left[(\mathbf{S}-\hat{\mathbf{S}})^2\right]$, must satisfy:
    \begin{itemize}
        \item Cell condition: Each quantization cell $\mathcal{R}_\ell$ $\;\ell=1,\ldots,L$, satisfies
        \begin{align}
             \mathcal{R}_{\ell} &= \left\{ \mathbf{x}: \norm{g(\mathbf{x})-\hat{\mathbf{s}}_{\ell}}^{2} \leq \norm{g(\mathbf{x})-\hat{\mathbf{s}}_{j}}^{2}, \, \forall j\neq \ell \right\},\label{eq:vec_thr_cond}
            \end{align}
            where
            $g(\mathbf{x})=\mathbb{E}[\mathbf{S} \mid \mathbf{X} = \mathbf{x})]$.
        \item Centroid condition:
         \begin{align}
            \hat{\mathbf{s}}_{\ell} &= \mathbb{E}\left[\mathbf{S}|\mathbf{X}\in \mathcal{R}_{\ell} \right],\;\ell=1,\ldots,L.\label{eq:value_cond_corr}
        \end{align}
    \end{itemize}
\end{claim}

\subsection{Necessary  Conditions for Optimality: Threshold Constraint}\label{sec:indirect_vector_thr}
\begin{figure}[ht]
        \centering
        \includegraphics[width=\columnwidth]     
        {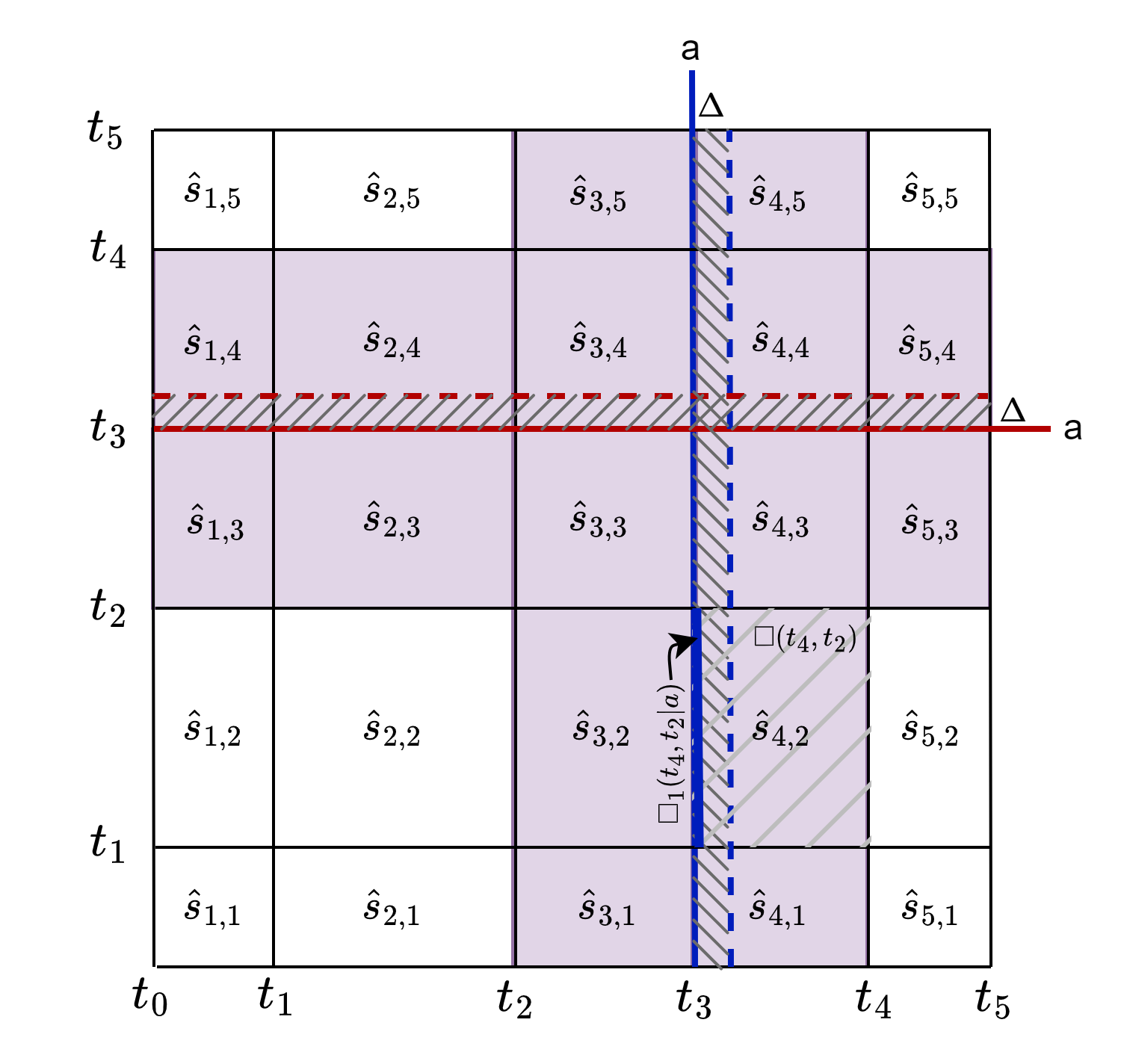}
        \caption{Necessary conditions for optimality for two-dimensional {($n=2$)} threshold-constrained quantization: 
        the shaded (purple) cells are the cells affected by adjusting a single threshold ($t_3$).}
        \label{fig:2D_grid}
    \end{figure}

We now present results for threshold-constrained indirect  quantization from a vector of observations.
The quantization cells must now be  hypercube as illustrated in Figure~\ref{fig:2D_grid} (for the two-dimensional case).

Rather than indexing the quantization cells by the index $\ell \in \{1,\ldots,L\}$, as was hitherto done, it will now be convenient to  index  quantization cells by an index vector 
$\mathbf{i} = (i_1,\ldots,i_n)$, where in each dimension $j\in\{1,\ldots,n\}$, the sub-index takes values 
 $i_j \in \{1,\ldots,T\}$.

The quantization cell corresponding to index $\mathbf{i}$ is defined (i.e., parameterized) by a vector of $n$ thresholds $\mathbf{t_{i}}$. Specifically, it is defined as the hypercube 
\begin{align}
\rect(\mathbf{t_{i}}) &= \rect\left([t_{i_1-1},t_{i_1}],\ldots,[t_{i_n-1},t_{i_n}]\right)
\nonumber
\\
&\triangleq 
[t_{i_1-1},t_{i_1}] \times \ldots \times [t_{i_n-1},t_{i_n}].
\label{eq:square_def}
\end{align}
We denote the reconstruction vector associated with cell $\mathbf{i}$, by $\mathbf{\hat{s}}_\mathbf{i} \in \mathbb{R}^k$.

We further denote  a face of a quantization cell, with respect to coordinate $j$,  by
\begin{align}
& \rect_j(\mathbf{t_{i}}| a) \nonumber \\
& = \rect\left([t_{i_1-1},t_{i_1}],\ldots,[t_{i_j -1} = a, t_{i_j }=a],\ldots,[t_{i_n-1},t_{i_n}] \right)
\label{eq:cond_square_def}
\end{align}


For brevity, we derive the necessary conditions for a scalar source $S$ and a two-dimensional observation vector $\mathbf{X} = [X_1,X_2]$. A  threshold-constrained vector is thus specified by a pair of threshold values $\mathbf{t_i} = (t_{i_1},t_{i_2})$.  The joint distribution of the pair $(S,\mathbf{X})$ is denoted by $f_{S\mathbf{X}}(s,\mathbf{x})$.

The definitions \eqref{eq:square_def}, 
\eqref{eq:cond_square_def} are illustrated in Figure~\ref{fig:2D_grid}.  The cell $\rect(t_4,t_2)$ ($i_1=4,i_2=2$) is the Cartesian product (rectangle) $[t_3,t_4] \times [t_1,t_2]$ and the associated reconstruction vector is denoted by $\hat{\mathbf{s}}_{4,2}$.
Note that only the upper thresholds (in this example $t_4$ and $t_2$) appear explicitly in the notation ($t_3$ and $t_1$ remain implicit).
The left face of the cell, marked by blue, is denoted by the interval  \mbox{$\rect_1(t_4,t_2|a)= [t_3 =a,t_4 =a] \times [t_1,t_2]$} and corresponds to  $\rect(t_4,t_2)$ where we set $t_3=t_4=a$, \emph{only} in dimension 1.


When deriving the necessary conditions for optimality of the thresholds, one needs to consider the effect of adjusting the value of a threshold $t_\ell$ on the MSE. This incremental change in the MSE is captured by the following two terms:
    \begin{align}
        &B_1(t_{\ell}) \triangleq  \sum_{i_2=1}^T \frac{1}{2}\left(\hat{s}_{_{\ell+1},{i_2}}^2 - \hat{s}_{_{\ell,i_2}}^2\right)
p_{\mathbf{X}}\left( \rect_1(\mathbf{t}_{i}|t_{\ell})\right)
\nonumber \\
        & + p_{\mathbf{X}}\left( \rect_1(\mathbf{t}_{i}|t_{\ell})\right) \left(\hat{s}_{_{\ell+1},i_2} - \hat{s}_{_{\ell,i_2}}\right) \mathbb{E}\left[S|  \mathbf{X} \in \rect_1(\mathbf{t}_{i}|t_{\ell})\right]\label{eq:B1_def}\\
        &B_2(t_{\ell}) \triangleq \sum_{i_1=1}^T \frac{1}{2}\left(\hat{s}_{_{i_1,\ell+1}}^2 - \hat{s}_{_{i_1,\ell}}^2\right)
        p_{\mathbf{X}}\left( \rect_2(\mathbf{t}_{i}|t_{\ell})\right)
\nonumber \\
        &+ p_{\mathbf{X}}\left( \rect_2(\mathbf{t}_{i}|t_{\ell})\right) \left(\hat{s}_{_{i_1,\ell+1}} - \hat{s}_{_{i_1,\ell}} \right) 
        \mathbb{E}\left[S|   \mathbf{X}\in\rect_2(\mathbf{t}_{i}|t_{\ell})\right] \label{eq:B2_def}
    \end{align}

where 
\begin{align}
  p_{\mathbf{X}}\left( \rect_1(\mathbf{t}_{i}|t_{\ell})\right) \triangleq \int_{t_{i_2-1}}^{t_{i_2}} f_{\mathbf{X}}(t_{\ell},x_2)dx_2 \nonumber \\
  =\frac{ \partial \Pr \left(\mathbf{X}\in \rect(t_{\ell},t_{i_2})\right)}{\partial t_{\ell} },
  \label{eq:half_prob1}
\end{align}
and 
\begin{align}
  p_{\mathbf{X}}\left( \rect_2(\mathbf{t}_{i}|t_{\ell})\right) \triangleq \int_{t_{i_1-1}}^{t_{i_1}} f_{\mathbf{X}}(x_1, t_{\ell})dx_1 \nonumber  \\
  =\frac{ \partial \Pr \left(\mathbf{X}\in \rect(t_{i_1},t_{\ell})\right)}{\partial t_{\ell} }.
    \label{eq:half_prob2}
\end{align}

These two terms may be understood by considering the effect of adjusting the value of $t_3$ in Figure~\ref{fig:2D_grid}.
Since we are assuming the same scalar quantizer is used over all dimensions, the set of cells affected is as illustrated in the figure as purple. The differential elements of the latter cells corresponding to an increment $\Delta$ appear as diagonally striped. For the cell $\rect(t_4,t_2)$, the probability of the observation vector falling in this striped (affected) region $\Delta \cdot p_{\mathbf{X}}\left( \rect_1(t_4,t_2|a)\right)$.
Finally, the change in MSE when perturbing the value of $t_3$ amounts to the sum of the term $\Delta \cdot B_1(t_\ell)$ (contribution from all rectangles belonging to the blue striped region) and $\Delta \cdot B_2(t_\ell)$ (contribution from red striped region), which should be stationary for an optimal set of thresholds. These considerations lead to the following set of necessary conditions for optimality.



\begin{lemma}[Indirect Two-Dimensional Quantization, Thresholds Constraint]\label{lem:indirect_2D}
Consider  indirect quantization of a scalar source $S$ from the observation $\mathbf{X}\in\mathbb{R}^2$, where the pair $(S,\mathbf{X})$ has joint PDF $f_{S\mathbf{X}}(s,\mathbf{x})$. An optimal quantizer, i.e., one that minimizes the MSE $\mathbb{E}\left[(S-\hat{S})\right]^2$, must satisfy:

\begin{itemize}
    \item Boundary condition: Each threshold $t_\ell$ 
    satisfies
        \begin{align}
            B_1(t_{\ell}) + B_2(t_{\ell}) = 0,\;\ell = 1,\ldots, L-1,\label{eq:bound_cond_2d}
        \end{align}
        where $t_0=-\infty$ and $t_L= \infty$, and $B_1(\cdot)$ and $B_2(\cdot)$ are defined in \eqref{eq:B1_def} and \eqref{eq:B2_def}, respectively.  
    \item Centroid condition:
       \begin{align}
            \hat{s}_{i_1,i_2} &= \mathbb{E}\left[ S|\mathbf{X}\in\rect(t_{i_1},t_{i_2})\right],\;i_1,i_2=1,\ldots,L,\label{eq:value_cond_2d}
        \end{align}
        where $\rect(t_{i},t_j)$ is defined in \eqref{eq:square_def}.
\end{itemize}

\begin{proof}
    The proof is given in the Appendix.
\end{proof}

\begin{remark}
     While the centroid condition is immediate, the boundary condition deserve discussion. 
 It is easiest to understand the condition and its derivation by comparison to its scalar counterpart appearing in Section~\ref{sec:extended_fine}. Specifically, compare Figure~\ref{fig:2D_grid} and Figure~\ref{fig:1D_grid}. While perturbation of a threshold affects only two cells in the scalar case, $4T$ cells in the two dimensional case (the shaded ``cross''). This in turn translates the boundary condition \eqref{eq:simple_boundary1} to the boundary condition \eqref{eq:bound_cond_2d} which involves a summation over all dimensions (two dimensions under the assumptions of this subsection). 
\end{remark}

\end{lemma}

\begin{remark}
It should be noted that in contrast to the scalar case, where  \eqref{eq:simple_boundary1} leads to the explicit condition \eqref{eq:thr_cond} the boundary condition \eqref{eq:bound_cond_2d} requires solving an implicit equation to find $t_{\ell}$, where all reconstruction points $\hat{s}_{i_1,i_2}$, and all other thresholds $t_j$, $j\neq \ell$, are held fixed. Furthermore, there might be several solutions for $t_{\ell}$ satisfying this condition. The centroid condition \eqref{eq:value_cond_2d} is the center of mass of each rectangle.
\end{remark}

\label{sec:s_vec}
\subsubsection{Multidimensional Vector Source}
The results are readily extended to a multidimensional vector source, namely $\mathbf{S}\in\mathbb{R}^K$, while retaining a two-dimensional observation vector. 
Apparently, the derivation is similar to the proof of Lemma \ref{lem:indirect_2D}, appearing in the Appendix, where one replaces the MSE as expressed in \eqref{eq:MSE_2D} with
\begin{align}
    \text{MSE} &= \sum_{i_1,\,i_2 } \int_\mathbf{S} \norm{\mathbf{s}-\hat{\mathbf{s}}_{_{i_1,i_2}}}^2 \int_{t_{i_1-1}}^{t_{i_1}}\int_{t_{i_2-1}}^{t_{i_2}}f_{\mathbf{SX}}(\mathbf{s},\mathbf{x})\mathbf{dx ds}.\label{eq:MSE_2D_s_vec}
\end{align}

The  boundary condition \eqref{eq:bound_cond_2d} and centroid condition \eqref{eq:value_cond_2d} become as follows: 
\begin{itemize}
    \item Boundary condition: 
        \begin{align}
            B_1(t_{\ell}) + B_2(t_{\ell}) = 0,\;\ell = 1,\ldots, L-1,\label{eq:bound_cond_2d_svec}
        \end{align}
        where $t_0=-\infty$ and $t_L= \infty$, and where 
    \begin{align}
        &B_1(t_{\ell}) \triangleq  \sum_{i_2=1}^T \frac{1}{2}\left(\norm{\hat{\mathbf{s}}_{_{\ell+1},{i_2}}}^2 - \norm{\hat{\mathbf{s}}_{_{\ell,i_2}}}^2\right) 
        p_{\mathbf{X}}\left(\rect_1(\mathbf{t}_{i}|t_{\ell})\right) \nonumber \\
        & + p_{\mathbf{X}}\left(\rect_1(\mathbf{t}_{i}|t_{\ell})\right) \left(\hat{\mathbf{s}}_{_{\ell+1},i_2} - \hat{\mathbf{s}}_{_{\ell,i_2}}\right)^T \mathbb{E}\left[\mathbf{S}|  \mathbf{X} \in \rect_1(\mathbf{t}_{i}|t_{\ell})\right]\label{eq:A1_def2}\\
        &B_2(t_{\ell}) \triangleq \sum_{i_1=1}^T \frac{1}{2}\left(\norm{\hat{\mathbf{s}}_{_{i_1,\ell+1}}}^2 - \norm{\hat{\mathbf{s}}_{_{i_1,\ell}}}^2\right)
         p_{\mathbf{X}}\left(\rect_2(\mathbf{t}_{i}|t_{\ell})\right)
        \nonumber\\
        &+ 
         p_{\mathbf{X}}\left(\rect_2(\mathbf{t}_{i}|t_{\ell})\right)
        \left(\hat{\mathbf{s}}_{_{i_1,\ell+1}} - \hat{\mathbf{s}}_{_{i_1,\ell}} \right)^T
        \mathbb{E}\left[\mathbf{S}|   \mathbf{X}\in\rect_2(\mathbf{t}_{i}|t_{\ell})\right] \label{eq:A2_def2}
    \end{align}
where $(\cdot)^T$ is the transpose operation, 
and where $p_{\mathbf{X}}\left(\rect_j(\mathbf{t}_{i}|t_{\ell})\right)$, $j=1,2$, are defined similarly to \eqref{eq:half_prob1} and \eqref{eq:half_prob2}.
    \item Centroid condition:
       \begin{align}
            \hat{\mathbf{s}}_{i_1,i_2} &= \mathbb{E}\left[\mathbf{S}|\mathbf{X}\in\rect(t_{i_1},t_{i_2})\right], \;i_1,i_2=1,\ldots,L.
            \label{eq:value_cond_svec}
        \end{align}
\end{itemize}
\subsubsection{Multidimensional Source and Observation Vectors}

When the dimension of the observation vector is also $n>2$, i.e., the most general case, 
Lemma~\ref{lem:indirect_2D} is readily extended using the following notation:
\textcolor{black}{
\begin{align}
&B_j(t_{\ell}) \triangleq  \sum_{\mathbf{i}_j} \frac{1}{2}\left(\norm{\hat{\mathbf{s}}_{_{\mathbf{i}_j(\ell+1)}}}^2 - \norm{\hat{\mathbf{s}}_{_{\mathbf{i}_j(\ell)}}}^2\right)
p_{\mathbf{X}}\left(\rect_j(\mathbf{t}_{i}|t_{\ell})\right)
\nonumber \\
& - p_{\mathbf{X}}\left(\rect_j(\mathbf{t}_{i}|t_{\ell})\right)
    \left(\hat{\mathbf{s}}_{_{\mathbf{i}_j(\ell+1)}} - \hat{\mathbf{s}}_{_{\mathbf{i}_j(\ell)}}\right)^T
    \mathbb{E}\left[\mathbf{S}| \mathbf{X} \in \rect_j(\mathbf{t}_{\mathbf{i}}|t_{\ell})\right]\label{eq:B_j},
\end{align}
where $\mathbf{i}_j \triangleq [i_1,\ldots,i_{j-1},i_{j+1},\ldots i_n]$, and $\mathbf{i}_j(\ell) \triangleq [i_1,\ldots,i_{j} = \ell,\ldots i_n]$,  
and where the $p_{\mathbf{X}}\left(\rect_j(\mathbf{t}_{i}|t_{\ell})\right)$ are defined similarly to \eqref{eq:half_prob1} and \eqref{eq:half_prob2}.
} 

The general vector case for indirect quantization  subject to a thresholds constraint is given in the following lemma:
\begin{lemma}[Indirect VQ , Thresholds Constraint]
\begin{itemize}
    \item Boundary condition: Each threshold $t_\ell$, $\;\ell=1,\ldots,L-1,$ satisfies
        \begin{align}
            \sum_{j=1}^{n}B_j(t_{\ell}) = 0,\;\ell = 1,\ldots, L-1\label{eq:indirect_vec_bound_cond}
        \end{align}
        where $t_0=-\infty$ and $t_L= \infty$.  
    \item Centroid condition:
       \begin{align}
            \mathbf{\hat{s}_i} &= \mathbb{E}\left(\mathbf{S}|\mathbf{X}\in\rect(\mathbf{t_{i}})\right),\;\mathbf{i}\in \{1,\ldots,L\}^n.\label{eq:indirect_vec_value_cond}
        \end{align}
\end{itemize}
\end{lemma}
\begin{remark}
    We note that these conditions may  be applied to yield an iterative quantizer design algorithm. Nonetheless, there is a freedom in choosing the order in which one proceeds between updating the threshold  and reconstruction values. 
\end{remark}
\begin{remark}\label{rem:memoryless}
    The terms $B_j(t_{\ell})$ for $j=1,\ldots,n$ in \eqref{eq:B_j} and fixed $t_{\ell}$ depend on $f_{\mathbf{X|S}}(\mathbf{x}|\mathbf{s})$ and $f_{\mathbf{S}}(\mathbf{s})$. For conditionally i.i.d. observations $X_1,\ldots,X_n$ given $\mathbf{S}$, i.e., $f_{\mathbf{X|S}}(\mathbf{x}|\mathbf{s}) = \prod_{i=1}^n f_{X|\mathbf{S}}(x_i|\mathbf{s})$, the summation inside  $B_j(t_{\ell})$ \eqref{eq:B_j} is identical for any permutation order of the $n$ dimensions. Consequently, we have that 
    $B_1(t_{\ell}) = B_2(t_{\ell}) = \cdots = B_n(t_{\ell})$, and the boundary condition \eqref{eq:indirect_vec_bound_cond} becomes
    \begin{align}
        B_1(t_{\ell}) = 0,\;\ell=1,\ldots,L-1.
    \end{align}
    The centroid condition \eqref{eq:indirect_vec_value_cond} satisfies $\mathbf{\hat{s}_i} = \mathbf{\hat{s}_{\pi(i)}}$, where $\mathbf{\pi(i)}$ is any permutation of  $\mathbf{i}$. Following the method of types \cite{CoverBook}, the reconstruction values depend on the type of hypercube to which $\mathbf{X}$ belongs. This property reduces the complexity to that of computing $\binom{n+L-1}{L-1}$ distinct reconstruction values (types), which is bounded by $(n+1)^{L-1}$.
   \end{remark}


\section{Numerical Results}\label{sec:example}
We  demonstrate the results via two examples.
The first example is the standard design problem of a direct scalar quantizer which we demonstrate for the case of a Gaussian source. 
The second example demonstrates the application of the derived results to the problem of parameter estimation based on quantized observations. This is illustrated via a Gaussian mixture model  
 . 



{\subsection{Direct Quantization: Scalar Gaussian Source}}\label{sec:example_direct_scalar}

Consider the following zero-mean Gaussian source with variance $\sigma_s^2$, i.e.,  
 \begin{align}
     f_{S}(s) &= \frac{1}{\sqrt{2\pi \sigma_s^2}} \exp\left\{-\frac{s^2}{2\sigma_s^2}\right\},
\label{eq:gaussian_source}
\end{align}
where $\sigma_s$ is a fixed and known standard deviation.

\begin{figure}[htbp]
        \centering
      \includegraphics[width=\columnwidth]{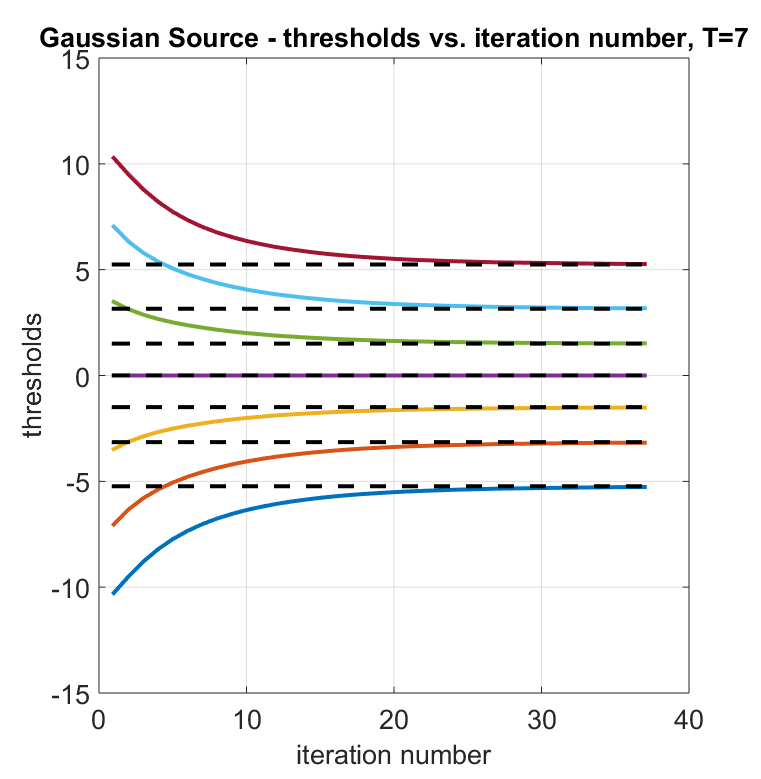}
        \caption{Gaussian source - evolution of threshold values as a function of the number of iterations of the Lloyd-Max algorithm (color) vs. optimal solution via Bruce's dynamic-programming algorithm (dashed) \cite{bruce1965optimum}; $T=7$ thresholds and $\sigma_s=3$. }
        \label{fig:thr_loc_T6_direct}
\end{figure}

The Lloyd-Max algorithm \cite{lloyd1982least,max1960quantizing} is known to convergence to the optimal scalar quantizer for log-concave PDFs \cite{fleisher1964sufficient}.  Figure~\ref{fig:thr_loc_T6_direct} demonstrates this convergence 
for the source \eqref{eq:gaussian_source} with $\sigma_s=3$ and 
$T=7$ thresholds. That is, it depicts the convergence of the  thresholds 
computed via the Lloyd-Max algorithm to the optimal thresholds, computed via the algorithm of Bruce \cite{bruce1965optimum} (described in Section~\ref{sec:direct_scalar}), as a function of the iteration number.

{\subsection{Indirect Quantization: 
Parameter Estimation from Quantized Samples of a Gaussian Mixture Model}}\label{sec:example_indirect_both}

\begin{figure}[ht]
    \centering
    \includegraphics[width=8.5cm,height=5.0cm]{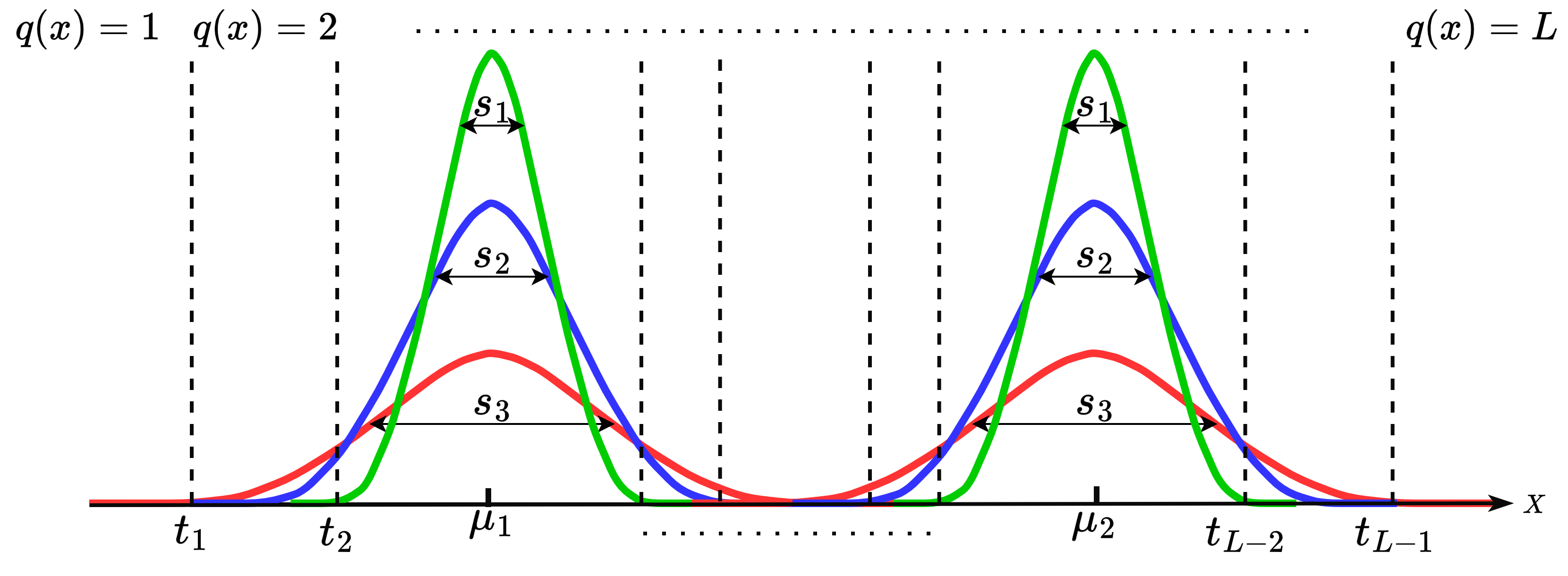}
    \caption{Bayesian parametric model: The observation  $X$ is drawn from a Gaussian mixture given the value of the source $S$.}
    \label{fig:model_example1}
\end{figure}

In this section we demonstrate the application of the indirect quantization framework to the problem of parameter estimation from quantized samples. 
To that end, 
assume the following Gaussian mixture model with two components of equal weight where the observation is distributed according to 
\begin{align}
    f_{X|S}(x|s) &= \sum_{i=1}^2 \frac{1}{2} \mathcal{N}(x|s,\mu_i),
    \label{eq:mixture_model1}
\end{align}
where
\begin{align}
    \mathcal{N}(x|s,\mu_i) &\triangleq \frac{1}{\sqrt{2\pi s^2}} \exp\left\{-\frac{(x-\mu_i)^2}{2s^2}\right\}, 
     \label{eq:mixture_model2}
\end{align}
and $\mu_1,\mu_2$ are known and deterministic. The model is illustrated in  Figure~\ref{fig:model_example1}.

Given $S=s$, the observation samples are drawn i.i.d. from the Gaussian mixture. The standard deviation $S$ of the Gaussian components is the hidden (unobserved) source.
We assume that $\mu_1 = -5$ and $\mu_2=5$. We further assume that $S$ is uniformly distributed over the interval $[1,2]$.


Our goal is to estimate the value of $S$ given quantized values of the observations $X_1,\ldots,X_n$. Thus, we wish to design the indirect quantizer, i.e., optimize the thresholds, so as to minimize the resulting MSE of the estimation of $S$. We first consider the case of a scalar observation ($n=1$) and then the case of an observation vector ($n\geq 2$). 

\subsubsection{Parameter Estimation from a Quantized Scalar Observation}
\label{sec:example_indirect_scalar}

Consider the case of observing a scalar observation, i.e., $n=1$. 

We  design the quantizer by the iterative Algorithm~\ref{algo:indirect_LM}, described in Section~\ref{sec:indirect_scalar_iter_thr_algo},
as well as by the dynamic-programming Algorithm~\ref{algo:indirect_scalar_thr_DP} described in Section~\ref{sec:indirect_scalar_algo_thr_opt}.

\begin{figure}[htbp]
        \centering
     \includegraphics[width=\columnwidth]{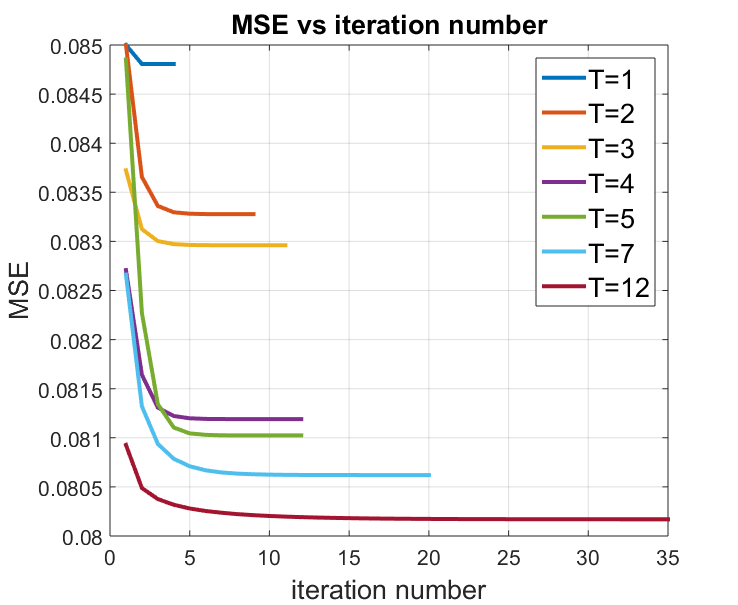}
        \caption{Gaussian mixture model \eqref{eq:mixture_model1}: MSE as a function of iteration number, when applying Algorithm~\ref{algo:indirect_LM} with $T=1,2,3,4,5,7,12$ thresholds. }
        \label{fig:MSE_vs_iter}
\end{figure}

In Figure~\ref{fig:MSE_vs_iter} we present the evolution of the MSE as the number of iterations grows, where the quantizer is iteratively updated according to Algorithm~\ref{algo:indirect_LM}. The results are presented for quantization with $T=1,2,3,4,5,7,12$ thresholds. 

\begin{figure}[htbp]
    \centering
    \subfloat[$T=2$ thresholds.]{\includegraphics[width=0.5\columnwidth, height=5.5cm]{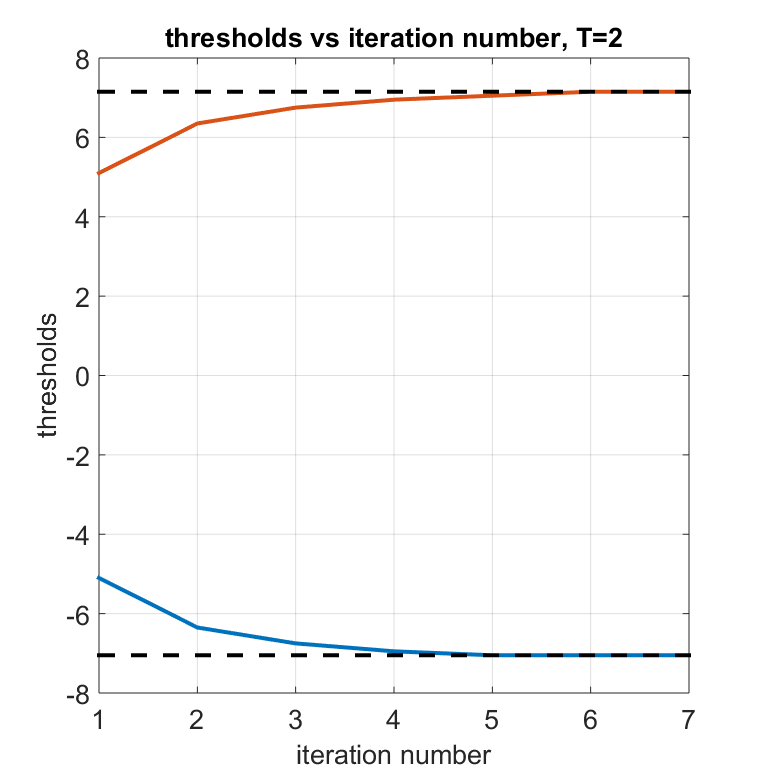} 
    \label{fig:thr_loc_T2}}
    \hfill
    \hspace{-0.6cm}
    \subfloat[$T=7$ thresholds.]{\includegraphics[width=0.5\columnwidth, height=5.5cm]{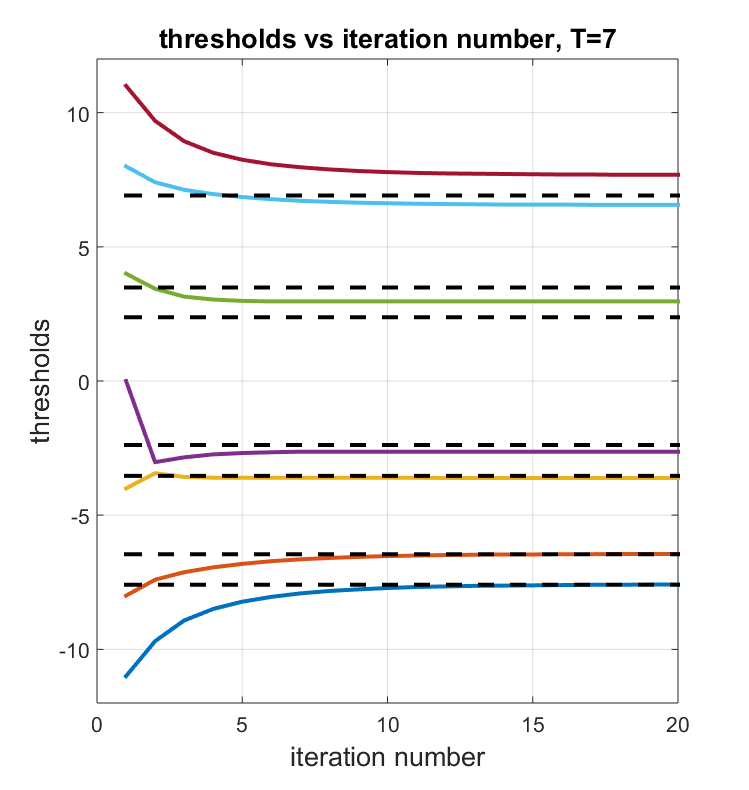} \label{fig:thr_loc_T7}}
    \caption{Gaussian mixture model \eqref{eq:mixture_model1}: Evolution of threshold values as a function of the number of iterations vs. optimal thresholds setting }
    \label{fig:thr_loc_T2_T7}
\end{figure}

The evolution of the threshold values as a function of the number of iterations is depicted in Figure~\ref{fig:thr_loc_T2} and Figure~\ref{fig:thr_loc_T7}, for $T=2$ and $T=7$ thresholds (color), respectively. The optimal threshold values, computed by Algorithm~\ref{algo:indirect_scalar_thr_DP}, are also depicted (dashed lines). Algorithm~\ref{algo:indirect_LM} is not guaranteed to converge to an optimal solution. This is illustrated in Figure~\ref{fig:thr_loc_T7} where, for $T=7$, the algorithm  
converges to a local minimum. On the other hand, in Figure~\ref{fig:thr_loc_T2}, for $T=2$, Algorithm~\ref{algo:indirect_LM} converges to the global minimum. 
In general, the outcome of the iterative algorithm depends on the initial conditions.

For  $T=2$ (two thresholds), the optimal threshold settings are symmetric around zero, since the Gaussian mixture model \eqref{eq:mixture_model1} is symmetric around $x=0$ when $\mu_1 = -\mu_2$. However, for $T=7$, an odd number of thresholds, the symmetric solution is disrupted by an additional threshold that is arbitrarily added to one of the sides.

\begin{remark}
Note that a rate-constrained quantizer would yield different results. Specifically, consider the case of $T=2$. In a rate-constrained scenario, the two extreme (most positive and most negative) cells would be mapped to a single index.    
\end{remark}

\begin{figure}[htbp]
        \centering
      \includegraphics[width=\columnwidth]{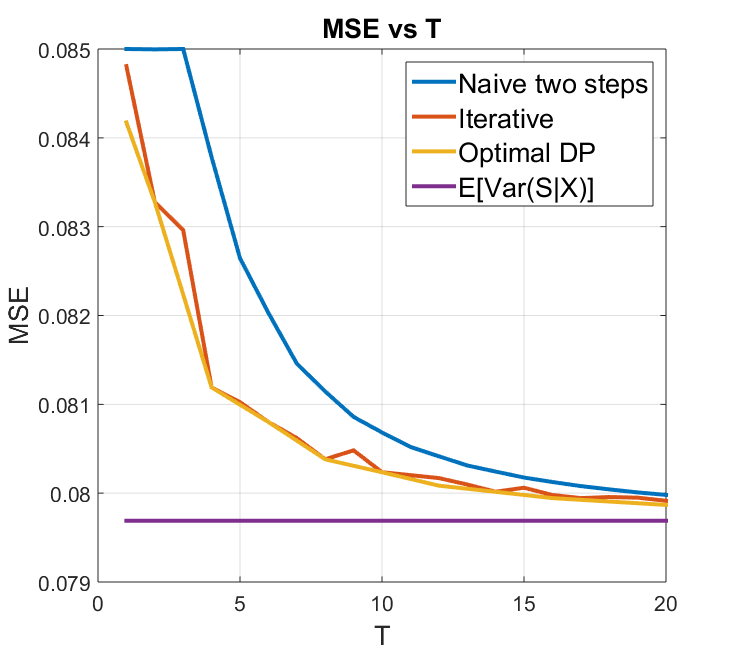}
        \caption{Gaussian mixture model \eqref{eq:mixture_model1}: MSE of indirect Lloyd-Max quantizer versus naive two-step quantizer as well as the optimal quantizer obtained via dynamic programming.}
   \label{fig:indirect_vs_two_phase}
\end{figure}
As a benchmark for comparison, we compare the obtained performance with that of the following two-step \emph{naive approach} to the indirect quantization problem: 
\begin{itemize}
    \item {Step I: Set threshold values via Bruce's algorithm \cite{bruce1965optimum} for direct quantization (Section~\ref{sec:direct_scalar}). That is, the thresholds minimize $\text{MSE}=\mathbb{E}\left[(X-\hat{X})^2\right]$.}  
    \item Step II: MMSE estimation of the latent variable $S$ given $q(X)$, the quantization of $X$, i.e.,
    \[\hat{S} = \mathbb{E} \left[S|q(X) \right].\]
\end{itemize}
We note that this naive solution is referred to as a ``task ignorant'' quantizer in \cite{shlezinger2019hardware}.


In Figure~\ref{fig:indirect_vs_two_phase} we compare the performance obtained via Algorithm~\ref{algo:indirect_LM} (iterative setting of thresholds), the  optimal indirect quantizer (Algorithm~\ref{algo:indirect_scalar_thr_DP}) and the naive solution. We demonstrate the gain in terms of the MSE, $\mathbb{E}\left[(S-\hat{S})^2\right]$, as a function of the number of thresholds $T$. Figure~\ref{fig:indirect_vs_two_phase} illustrates that for the optimal solution of Algorithm~\ref{algo:indirect_scalar_thr_DP}, the MSE monotonically decreases as the number of thresholds $T$ increases. In contrast, the iterative solution does not guarantee \textit{strict} monotonicity, as the attained MSE depends on the initial conditions. 

\begin{remark}
As another baseline, we note that the MSE is also bounded from below by the conditional variance of the source $S$ given the unquantized observation (which corresponds to $ T\rightarrow\infty$), i.e., by $\mathbb{E}[\text{Var}(S|X)]$  as shown in Figure~\ref{fig:indirect_vs_two_phase}. Indeed, as the number of thresholds increases, the improvement in MSE (of the estimators based on a quantized observation) diminishes as one approaches this asymptotic MSE.
\end{remark}



\subsubsection{Parameter Estimation from a Quantized Observation Vector}\label{sec:example_indirect_vector}


We now consider the Gaussian mixture model \eqref{eq:mixture_model1}, \eqref{eq:mixture_model2},  
for the case of a vector of observations, i.e., $n \geq 2$. Specifically, we assume that, given $S$, the observations
$\mathbf{X}$ are i.i.d. are distributed according to
\begin{align}
    f_{\mathbf{X}|S}(\mathbf{x}|s) &= \prod_{j=1}^n f_{X|S}(x_j|s),\label{eq:mixture_vector_model}
\end{align}
where $f_{X|S}(x|s)$ is given by \eqref{eq:mixture_model1}.

We demonstrate the iterative algorithm developed in Section~\ref{sec:indirect_vector_thr} for the design of a threshold-constrained indirect scalar quantizer for the quantization of a vector of such observations.\footnote{We note that unlike for the case of a scalar observation, considered in the previous subsection, a dynamic-programming solution is not known for the vector case.}
We first present the results and then give some details related to the algorithm implementation.

\begin{figure}[htbp]
        \centering
      \includegraphics[width=\columnwidth]{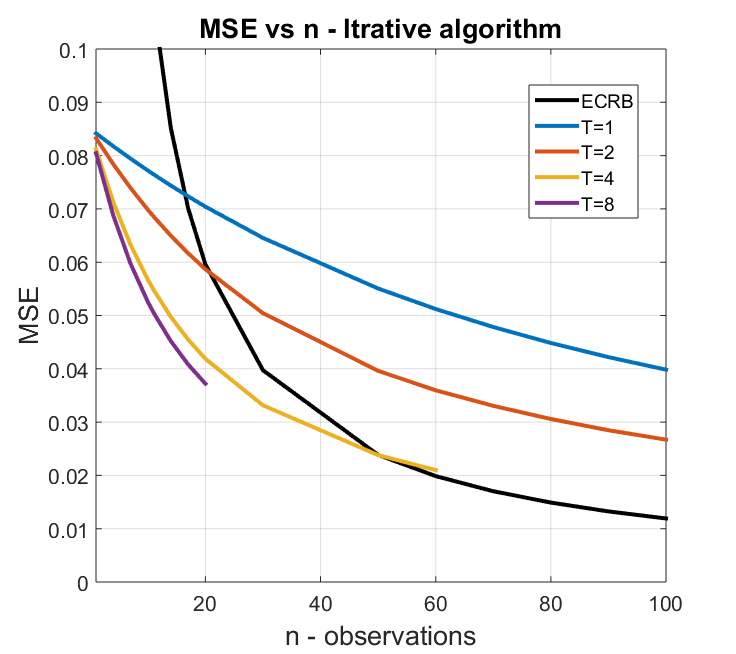}
        \caption{
    Gaussian mixture model \eqref{eq:mixture_model1}: MSE of the estimation $\hat{S}$ versus number of quantized observations $n$, for a various number of thresholds $T=1, 2, 4, 8$. The thresholds are set using the iterative algorithm described in Section~\ref{sec:indirect_vector_thr}.}
   \label{fig:indirect_vector}
\end{figure}

Figure~\ref{fig:indirect_vector} depicts the obtained MSE versus the number of observations $n$ for a various number of thresholds $T$. Specifically, the MSE of the quantizer obtained using the iterative algorithm, for $T=1,2,4,8$, is shown by the colored curves. 
As a benchmark for comparison, we also depict the expected Cram\'{e}r-Rao bound (ECRB) \cite{aharon2024asymptotically}, computed for the unquantized output (which corresponds to $T\rightarrow\infty$).
The latter is a tight approximation for the optimal MSE (for unquantized observations) for large $n$ \cite{aharon2024asymptotically}. 
We note that, as expected, the MSE of all quantizers 
decreases at a rate of $1/n$ as $n$ grows.

Interestingly, the optimal indirect quantizer for $n=1$,
shows promising performance across dimensions as compared to the performance of the quantizer designed using the iterative algorithm.

In the remainder of this section we give some details concerning the implementation of the algorithm. 
We use the notation for the vector $\mathbf{x}$, interchangeably as $x_1,x_2\dots,x_n$ or $x_1^n$ when considering it as a series.

As noted in Remark~\ref{rem:memoryless}, when the observations are i.i.d. given $S$, the boundary and centroid conditions become:
\begin{itemize}
    \item Boundary condition: 
        \begin{align}
            B_1(t_{\ell}) = 0,\;\ell = 1,\ldots, L-1,
        \end{align}
        where $t_0=-\infty$ and $t_L= \infty$ and $B_1(t_{\ell})$ is defined in \eqref{eq:B_j} for $j=1$.
    \item Centroid condition:
       \begin{align}
            \hat{s}_i &= \mathbb{E}\left[S|\mathbf{X}\in\rect(\mathbf{t_{i}})\right],\;\mathbf{i}\in \{1,\ldots,L\}^n.
        \end{align}
\end{itemize}

Thus, in order to carry out the iterative algorithm, we need to evaluate  $\mathbb{E}\left[S|\mathbf{X}\in\rect(\mathbf{t_{i}})\right]$ and $B_1(t_{\ell})$. These terms can be evaluated with polynomial complexity in $n$ using the method of types; see Section~12.1 in \cite{CoverBook}.

The sequence of quantized values $q(X_1), q(X_2),\ldots, q(X_n)$ are i.i.d. given $S=s$,  with probability mass function that we denote by $P(\cdot|s)$. Thus, {$P^n(q(x_1^n)|s) \triangleq \prod_{i=1}^n P(q(x_i)|s)$} denotes the probability of observing  $q(x_1),q(x_2),\ldots, q(x_n)$ given $S=s$. A type $Q$ of a sequence $q(x_1), q(x_1),\ldots, q(x_n)$ is the relative proportion of occurrences of each symbol in the vector $q(x_1^n) \in \{1,\ldots,L\}^n$. In other words, it's the empirical probability distribution.
The set of sequences of length $n$ and type $Q$
is called the type class of $Q$, denoted by $T(Q)$. We also denote the set of all types of ${q(x_1^n)}$ by $\mathcal{P}_n$, thus

\begin{align}
    \mathcal{P}_n = \left\{\frac{n_1}{n},\frac{n_2}{n},\ldots,\frac{n_L}{n}: n_i\in \mathbb{Z}^+,\; \sum_{i=1}^L n_i = n\right\},
\end{align}
where $\mathbb{Z}^+=\{0,1,2,\ldots\}$ are the non-negative integers.

We quote here only the final expressions, based on types, that are required to be evaluated in the iterative algorithm:
\begin{align}
& B_1(t_{\ell}) = \nonumber \\
&\sum_{Q'\in \mathcal{P}_{n-1}} 
|T(Q')|\cdot p_{\mathbf{X}}\left( x_1 = t_{\ell}, q(x_2^n) \in Q'\right)\nonumber \\
&\quad\left(\hat{s}_{_{[\ell+1,Q']} }- \hat{s}_{_{[\ell,Q']}}\right)
\bigg\{\frac{\left(\hat{s}_{_{[\ell+1,Q']}} + \hat{s}_{_{[\ell,Q']}}\right)}{2}
\bigg. \nonumber \\ 
& \bigg. 
    \hspace{3.3cm}-\mathbb{E}\left[S| X_1 = t_{\ell}, q(X_2^n) \in Q'\right]\bigg\}\\
    &\hat{s}_{_Q} =\mathbb{E}\left(S|q(\mathbf{X})\in Q)\right)
    = \frac{\int_{s \in \mathbb{R}} sf_{S}(s) P^{n}\left(q(x_1^{n} 
 ) \in Q |s\right) ds}{\Pr\left( q(\mathbf{X})\in Q\right)},
\end{align}
where $\mathcal{P}_{n-1}$ is the set of all types of sequences of length $n-1$, and $|T({Q'})|$ is the size of type class $Q'$ of length $n-1$. We denote by $\hat{s}_{_{[\ell,Q']}}$  the reconstruction value of the associated type in $\mathcal{P}_n$, where the first dimension is restricted to the symbol $\ell$ and the rest of the dimensions follow the occurrence of the type $Q'\in\mathcal{P}_{n-1}$. Further, 
\begin{align}
  &p_{\mathbf{X}}\left(  x_1 = t_{\ell}, q(x_2^n) \in Q'\right) \triangleq \nonumber \\
  &\qquad\qquad \int_{s \in \mathbb{R}} f_S(s) f_{X|S}(t_{\ell}|s) P^{n-1}\left(q(x_2^{n}) \in Q'|s\right)ds\\
    & \mathbb{E}\left[S| X_1 = t_{\ell}, q(X_2^n) \in Q'\right] = \nonumber \\
    &\qquad\qquad \frac{\int_{s \in \mathbb{R}} sf_S(s) f_{X|S}(t_{\ell}|s) P^{n-1}\left(q(x_2^{n}) \in Q'|s\right) ds}{p_{\mathbf{X}}\left( x_1 = t_{\ell}, q(x_2^n) \in Q'\right)}\\
    &\Pr\left( q(\mathbf{X})\in Q\right) =\int_{s \in \mathbb{R}} f_{S}(s) P^{n}\left(q(x_1^{n}) \in Q |s\right) ds.
\end{align}




\section{Discussion}
\label{sec:discussion}

We have studied the problem of indirect quantization under MSE distortion, where time-invariant threshold-constrained scalar quantization  is performed. Unlike in the case of direct quantization, where it is immaterial whether one imposes a  constraint on the number of thresholds used or on the rate of quantization, we have seen that for indirect quantization, the distinction between the two constraints is of critical importance. Necessary conditions for optimality, and associated iterative algorithms for the design of indirect quantizers were derived for both types of constraints.

We have also shown that the problem of Bayesian estimation of parameters from quantized samples can also be formulated under the framework of indirect quantization, as was illustrated by a numerical example. 

Whereas for the case of a one-dimensional observation, the optimal quantizer can be found  efficiently via a dynamic-programming algorithm, in the case of a multi-dimensional observation vector, we have relied on suboptimal iterative quantizer design algorithms. The development of efficient algorithms for the design of optimal scalar quantizers in the scenario of a multi-dimensional observation vector remains an open question for future research.

\appendices
\section{Proof of Lemma~\ref{lem:indirect_2D} }
We now prove Lemma~\ref{lem:indirect_2D}.
For the case of a scalar source and a two-dimensional observation vector, the MSE \eqref{eq:MSE_VQ_general} can be expressed explicitly as:
\begin{align}
   & \text{MSE} = \sum_{i_1,\,i_2 } \int_{s \in \mathbb{R}} (s-\hat{s}_{_{i_1,i_2}})^2 \int_{t_{i_1-1}}^{t_{i_1}}\int_{t_{i_2-1}}^{t_{i_2}}f_{S\mathbf{X}}(s,\mathbf{x})\mathbf{dx}\;ds ,
    \label{eq:MSE_2D}
\end{align}
where $\mathbf{x}\triangleq (x_1,x_2)$.
We evaluate the partial derivative of $\text{MSE}$ with respect to $t_{\ell}$, namely $\frac{\partial \text{MSE}}{\partial t_{\ell}}$. In the summation  \eqref{eq:MSE_2D}, there are five types of terms that depend on $t_{\ell}$; see Figure~\ref{fig:2D_grid}. Applying the Leibniz integral rule for double integrals, the partial derivatives of the inner double integral for these cases are:
\begin{align}
    &\frac{\partial}{\partial t_{\ell}}\int_{t_{\ell-1}}^{t_{\ell}}\int_{t_{i_2-1}}^{t_{i_2}} f_{S\mathbf{X}}(s,\mathbf{x})\mathbf{dx} = \int_{t_{i_2-1}}^{t_{i_2}} f_{S\mathbf{X}}(s,(t_{\ell},x_2))dx_2\label{eq:Leib_rule1}\\
    &\frac{\partial}{\partial t_{\ell}}\int_{t_{\ell}}^{t_{\ell+1}}\int_{t_{i_2-1}}^{t_{i_2}} f_{S\mathbf{X}}(s,\mathbf{x})\mathbf{dx} = -\int_{t_{i_2-1}}^{t_{i_2}} f_{S\mathbf{X}}(s,(t_{\ell},x_2)) dx_2\label{eq:Leib_rule2}\\
    &\frac{\partial}{\partial t_{\ell}}\int_{t_{\ell-1}}^{t_{\ell}}\int_{t_{\ell-1}}^{t_{\ell}} f_{S\mathbf{X}}(s,\mathbf{x})\mathbf{dx} = \int_{t_{\ell-1}}^{t_{\ell}} f_{S\mathbf{X}}(s,(x_1, t_{\ell}))dx_1  \nonumber \\ & \qquad +\int_{t_{\ell-1}}^{t_{\ell}} f_{S\mathbf{X}}(s,(t_{\ell},x_2)) dx_2 \label{eq:Leib_rule3}\\
    &\frac{\partial}{\partial t_{\ell}}\int_{t_{\ell}}^{t_{\ell+1}}\int_{t_{\ell}}^{t_{\ell+1}} f_{S\mathbf{X}}(s,\mathbf{x})\mathbf{dx} = -\int_{t_{\ell}}^{t_{\ell+1}} f_{S\mathbf{X}}(s,(x_1, t_{\ell}))dx_1  \nonumber \\ &\qquad  - \int_{t_{\ell}}^{t_{\ell+1}} f_{S\mathbf{X}}(s,(t_{\ell},x_2)) dx_2 \label{eq:Leib_rule4}\\
    &\frac{\partial}{\partial t_{\ell}}\int_{t_{\ell-1}}^{t_{\ell}}\int_{t_{\ell}}^{t_{\ell+1}} f_{S\mathbf{X}}(s,\mathbf{x})\mathbf{dx} = -\int_{t_{\ell}}^{t_{\ell+1}} f_{S\mathbf{X}}(s,(x_1, t_{\ell}))dx_1  \nonumber \\ &\qquad  + \int_{t_{\ell-1}}^{t_{\ell}} f_{S\mathbf{X}}(s,(t_{\ell},x_2)) dx_2\label{eq:Leib_rule5}.
\end{align}
In the same manner, the counterpart of \eqref{eq:Leib_rule1}-\eqref{eq:Leib_rule5}, where we swap the limits of the two integrals (on the left hand side), is swapping  the roles of $x_1$  and $x_2$  on the right hand sides of these equations. 
Consequently, applying properties \eqref{eq:Leib_rule1}-\eqref{eq:Leib_rule5} and their dual counterparts, we obtain that the derivative of the MSE \eqref{eq:MSE_2D}  is given by:
\begin{align}
& \frac{\partial \text{MSE}}{\partial t_{\ell}} =  \sum_{i_1=1}^T \int_{s \in \mathbb{R}} (s-\hat{s}_{_{i_1,\ell}})^2 \int_{t_{i_1-1}}^{t_{i_1}} f_{S\mathbf{X}}(s,(x_1,t_{\ell}))dx_1\;ds \nonumber \\
&\qquad - \int_{s \in \mathbb{R}} (s-\hat{s}_{_{i_1,\ell+1}})^2 \int_{t_{i_1-1}}^{t_{i_1}} f_{S\mathbf{X}}(s,(x_1,t_{\ell}))dx_1\;ds \nonumber \\
&\qquad +  \sum_{i_2=1}^T \int_{s \in \mathbb{R}} (s-\hat{s}_{_{\ell,i_2}})^2 \int_{t_{i_2-1}}^{t_{i_2}} f_{S\mathbf{X}}(s,(t_{\ell},x_2))dx_2\;ds \nonumber \\
&\qquad - \int_{s \in \mathbb{R}} (s-\hat{s}_{_{\ell+1,i_2}})^2 \int_{t_{i_2-1}}^{t_{i_2}} f_{S\mathbf{X}}(s,(t_{\ell},x_2))dx_2\;ds 
\end{align}
The last term can be rearranged as follows:
\begin{align}
& \frac{\partial \text{MSE}}{\partial t_{\ell}} =  \sum_{i_1=1}^T (\hat{s}_{_{i_1,\ell+1}}^2-\hat{s}_{_{i_1,\ell}}^2) \int_{t_{i_1-1}}^{t_{i_1}}  f_{\mathbf{X}}(x_1,t_{\ell})dx_1 \nonumber \\
&\quad + 2(\hat{s}_{_{i_1,\ell+1}}-\hat{s}_{_{i_1,\ell}}) \int_{s \in \mathbb{R}} \int_{t_{i_1-1}}^{t_{i_1}}  s f_{S\mathbf{X}}(s,(x_1,t_{\ell}))dx_1ds \nonumber \\
&\quad +  \sum_{i_2=1}^T  (\hat{s}_{_{\ell+1,i_2}}^2-\hat{s}_{_{\ell,i_2}}^2) \int_{t_{i_2-1}}^{t_{i_2}} f_{\mathbf{X}}(t_{\ell},x_2)dx_2  \nonumber \\
&\quad +  2(\hat{s}_{_{\ell+1,i_2}}-\hat{s}_{_{\ell,i_2}})\int_{s \in \mathbb{R}}  \int_{t_{i_2-1}}^{t_{i_2}} sf_{S\mathbf{X}}(s,(t_{\ell},x_2))dx_2ds\\
& = 2B_1(t_{\ell}) + 2B_2(t_{\ell}),
\end{align}
where the last equality follows \eqref{eq:half_prob1} and \eqref{eq:half_prob2}, thus 
\begin{align}
&\int_{t_{i_1-1}}^{t_{i_1}} f_{\mathbf{X}}(x_1,t_{\ell})dx_1 = p_{\mathbf{X}}\left( \rect_2(\mathbf{t}_{i}|t_{\ell})\right)\\
&\int_{t_{i_2-1}}^{t_{i_2}} f_{\mathbf{X}}(t_{\ell},x_2)dx_2 = p_{\mathbf{X}}\left( \rect_1(\mathbf{t}_{i}|t_{\ell})\right)\\
&\int_{t_{i_1-1}}^{t_{i_1}}  f_{S\mathbf{X}}(s,(x_1,t_{\ell}))dx_1 = \nonumber\\
&\quad p_{\mathbf{X}}\left( \rect_2(\mathbf{t}_{i}|t_{\ell})\right) f_{S}(s|\mathbf{X}\in  \rect_2(t_{i_1},t_{i_2}|t_{\ell}) \\
&\int_{t_{i_1-1}}^{t_{i_1}}  f_{S\mathbf{X}}(s,(t_{\ell}),x_2)dx_2 = \nonumber \\ 
&\quad p_{\mathbf{X}}\left( \rect_1(\mathbf{t}_{i}|t_{\ell})\right) f_{S}(s|\mathbf{X}\in  \rect_1(t_{i_1},t_{i_2}|t_{\ell}), 
\end{align}
and using the definition of conditional mean, and further applying the definitions of $B_1(\cdot)$ and $B_2(\cdot)$ in \eqref{eq:B1_def} and \eqref{eq:B2_def}. The boundary condition follows by requiring that $\frac{\partial \text{MSE}}{\partial t_{\ell}} =0$.

\bibliographystyle{IEEEtran}
\bibliography{mybib}

\end{document}